\documentclass[a4paper,11pt,centertags,psamsfonts]{amsart}
\usepackage{times,epsfig,amssymb,pifont}
\usepackage[mathcal]{euscript}

\DeclareSymbolFont{operators}   {OT1}{ptmcm}{m}{n}
\DeclareSymbolFont{letters}     {OML}{ptmcm}{m}{it}
\DeclareSymbolFont{symbols}     {OMS}{pzccm}{m}{n}
\DeclareSymbolFont{largesymbols}{OMX}{psycm}{m}{n}

\DeclareMathAlphabet{\mathsf}{OT1}{phv}{m}{n}
\DeclareMathAlphabet{\mathrm}{OT1}{ptm}{m}{n}

\DeclareSymbolFont{ER}{U}{eur}{m}{n}
\DeclareSymbolFont{SY}{U}{psy}{m}{n}

\DeclareMathSymbol{,}{\mathpunct}{SY}{'054}
\DeclareMathSymbol{.}{\mathpunct}{SY}{'056}
\DeclareMathSymbol{:}{\mathpunct}{SY}{'072}

\DeclareMathSymbol{(}{\mathopen}{SY}{'050}
\DeclareMathSymbol{)}{\mathclose}{SY}{'051}
\DeclareMathSymbol{[}{\mathopen}{SY}{'133}
\DeclareMathSymbol{]}{\mathclose}{SY}{'135}

\DeclareMathSymbol{+}{\mathbin}{SY}{'053}
\DeclareMathSymbol{-}{\mathbin}{SY}{'055}
\DeclareMathSymbol{=}{\mathbin}{SY}{'075}
\DeclareMathSymbol{<}{\mathbin}{SY}{'074}
\DeclareMathSymbol{>}{\mathbin}{SY}{'076}
\DeclareMathSymbol{\leq}{\mathbin}{SY}{'243}
\DeclareMathSymbol{\geq}{\mathbin}{SY}{'263}
\DeclareMathSymbol{\nneq}{\mathbin}{SY}{'271}
\DeclareMathSymbol{\nnotin}{\mathbin}{SY}{'317}
\DeclareMathSymbol{\in}{\mathbin}{SY}{'316}
\DeclareMathSymbol{\times}{\mathbin}{SY}{'264}
\DeclareMathSymbol{\pm}{\mathbin}{SY}{'261}
\DeclareMathSymbol{\subset}{\mathbin}{SY}{'314}
\DeclareMathSymbol{\supset}{\mathbin}{SY}{'311}
\DeclareMathSymbol{\subseteq}{\mathbin}{SY}{'315}
\DeclareMathSymbol{\supseteq}{\mathbin}{SY}{'312}

\DeclareMathSymbol{/}{\mathord}{SY}{'057}
\DeclareMathSymbol{\ast}{\mathord}{SY}{'052}
\DeclareMathSymbol{\perp}{\mathord}{SY}{'136}
\DeclareMathSymbol{\emptyset}{\mathord}{SY}{'306}
\DeclareMathSymbol{\oplus}{\mathord}{SY}{'305}

\renewcommand{\neq}{\nneq}
\renewcommand{\notin}{\nnotin}

\renewcommand{\theequation}{\arabic{section}.\arabic{equation}}

\newcommand{\R}{\mathbb{R}}
\newcommand{\C}{\mathbb{C}}

\newcommand{\E}{\mathbb{E}}

\newcommand{\cB}{\mathcal{B}}
\newcommand{\cC}{\mathcal{C}}
\newcommand{\cD}{\mathcal{D}}
\newcommand{\cE}{\mathcal{E}}

\newcommand{\cH}{\mathcal{H}}
\newcommand{\cI}{\mathcal{I}}
\newcommand{\cJ}{\mathcal{J}}

\newcommand{\cL}{\mathcal{L}}

\newcommand{\supp}{{\ensuremath{\mathrm{supp}}}}

\newcommand{\diag}{{\ensuremath{\mathrm{diag}}}}

\newcommand{\U}{\mathsf{U}}
\newcommand{\SL}{\mathsf{SL}}

\newcommand{\SU}{\mathsf{SU}}

\newcommand{\fH}{\mathfrak{H}}
\newcommand{\fN}{\mathfrak{N}}

\DeclareMathOperator*{\slim}{s-lim}
\DeclareMathOperator{\Ran}{\mathrm{Ran}}
\DeclareMathOperator{\Ker}{\mathrm{Ker}}
\DeclareMathOperator{\Rank}{\mathrm{Rank}}
\newcommand{\1}{\mathbb{I}}

\renewcommand{\det}{\mathrm{det\ }}

\newcommand{\linspan}{\mathrm{lin\ span}}

\newtheorem{theorem}{Theorem}[section]{\bf}{\it}
{\bf}{\it}
\newtheorem{corollary}[theorem]{Corollary}{\bf}{\it}
\newtheorem{example}[theorem]{Example}{\it}{\rm}
\newtheorem{lemma}[theorem]{Lemma}{\bf}{\it}
{\it}{\rm}
\newtheorem{definition}[theorem]{Definition}{\bf}{\it}

\title[Scattering Matrices
on Graphs]{The Generalized Star Product and the Factorization of Scattering
Matrices on Graphs}

\author[V. Kostrykin and R. Schrader]{V. Kostrykin \and R. Schrader$^\ast$}

\address{Vadim Kostrykin\\
Fraunhofer-Institut f\"{u}r Lasertechnik\\ Steinbachstra{\ss}e 15, D-52074\\ Aachen,
Germany}
\email{kostrykin@t-online.de, kostrykin@ilt.fhg.de}

\address{Robert Schrader\\ Institut f\"{u}r
Theoretische Physik\\ Freie Universit\"{a}t Berlin, Arnimallee 14\\ D-14195 Berlin,
Germany}
\email{schrader@physik.fu-berlin.de}

 \thanks{$^\ast$ R.S. supported in part by
DFG SFB 288 ``Differentialgeometrie und Quantenphysik''}

\date{August 14, 2000.}



\begin{document}

\begin{abstract}
In this article we continue our analysis of Schr\"{o}dinger operators on arbitrary
graphs given as certain Laplace operators. In the present paper we give the
proof of the composition rule for the scattering matrices. This composition
rule gives the scattering matrix of a graph as a generalized star product of
the scattering matrices corresponding to its subgraphs. We perform a detailed
analysis of the generalized star product for arbitrary unitary matrices. The
relation to the theory of transfer matrices is also discussed.
\end{abstract}

\maketitle
\thispagestyle{empty}

\section{Introduction}

Potential scattering for one particle Schr\"{o}dinger operators on the line
possesses a remarkable property concerning its (on-shell) scattering matrix
given as a $2\times 2$ matrix-valued function of the energy. Let the potential
$V$ be given as the sum of two potentials $V_{1}$ and $V_{2}$ with disjoint
support. Then the scattering matrix for $V$ at a given energy is obtained from
the two scattering matrices for $V_1$ and $V_2$ at the same energy by a certain
non-linear, noncommutative but associative composition rule. This fact has been
discovered independently by several authors (see e.g.\
\cite{Aktosun,Redheffer:61,Redheffer:62,Kowal,Rozman,Bianchi:Ventra:95,Bianchi:Ventra:95a})
and is an easy consequence of the multiplicative property of the transfer
matrix of the Schr\"{o}dinger equation (see e.g.\ \cite{Kostrykin:Schrader:99a}).
It has also been well known in the theory of mesoscopic systems and
multichannel conductors (see e.g.\ \cite{Tong,Dorokhov:82, Dorokhov:83,
Dorokhov:84,
Dorokhov:88,Mello:Pereyra:Kumar,Stone:Mello:Muttalib:Pichard,Beenakker,Datta,Exner:Tater:98}).
In higher space dimensions a similar rule is not known. However, for large
separation between the supports of the potentials the scattering matrix at a
given energy is asymptotically related to the scattering matrices for $V_1$ and
$V_2$ at the same energy \cite{Kostrykin:Schrader:94,Kostrykin:Schrader:98}.

To the best of our knowledge the composition rule for $2\times 2$ scattering
matrices was first observed in network theory by Redheffer
\cite{Redheffer:61,Redheffer:62}, who called it the star product. In our
preceding article \cite{Kostrykin:Schrader:99b} we extended this result to
quasi-one dimensional quantum systems -- Schr\"{o}dinger operators on graphs. Such
systems are nowadays a subject of intensive study (see e.g.\
\cite{Gerasimenko:Pavlov,Avron:Exner:Last,Exner:96,Carlson:97,Carlson:98}).
Some other related works are quoted in \cite{Kostrykin:Schrader:99b}. In
\cite{Rubinstein:Schatzman,Kuchment:Zeng,Saito,Saito:2000} differential
operators with Neuman boundary conditions on ``fat graphs" were considered,
i.e.\ on thin domains in $\R^d$ which asymptotically shrink to a graph.

There is also a large amount of literature on linear difference operators on
graphs. The motivation for the study of such operators comes from the graph
theory, where the spectrum of these operators are known to be related to
topological properties of the graph \cite{Cvetcovic:Doob:Sachs,Chung,Bollobas}.
Scattering theory for such operators was developed in \cite{Novikov:98,
Allard:Froese}.

In \cite{Kostrykin:Schrader:99b} we considered the (continuous) Laplace
operator on graphs with an arbitrary number $n$ of open ends (i.e.\ channels)
and with arbitrary boundary conditions at the edges resulting in a self-adjoint
operator. We formulated and proved necessary and sufficient conditions for such
operators to be self-adjoint. We provided an explicit expression for the
resulting unitary $n\times n$ scattering matrix in terms of the boundary
conditions, the lengths of the internal lines and the given energy.
Furthermore, we generalized Redheffer's star product to what we called the
generalized star product. This is a non-linear, noncommutative but associative
composition rule for unitary matrices not necessarily of equal rank and
resulting in a unitary matrix.

Under special circumstances there is an alternative way to describe the
generalized star product. Fix $p\geq 1$ and consider the group $\U(p,p)$ with
its natural multiplication. As a set this group is isomorphic to some subgroup
of $\U(2p)$. This non-linear set isomorphism is well known in the case $p=1$
(see e.g.\ \cite{Dubrovin:Novikov:Fomenko}) and can be easily generalized to
the case of arbitrary $p>1$. Under this isomorphism the multiplication in
$\U(p,p)$ induces new nonlinear multiplication $*_p$ in this subgroup of
$\U(2p)$, which is our generalized star product. The operation $*_p$ can be
extended by continuity to the whole $\U(2p)$. The set $\U(2p)$ with $*_p$ as
multiplication is no longer a group, but only a semigroup.

Employing this generalized star product in \cite{Kostrykin:Schrader:99b} we
provided a formal proof based on the quantum mechanical superposition principle
to show how the scattering matrix at the same energy for the whole graph can be
obtained from the scattering matrices of two subgraphs obtained by cutting the
graph in any way in two. Again for the special case of 2-channel scattering
matrices, like potential scattering on the line, this formal argument is well
known (see e.g.\ \cite{Datta}). In this article we will provide a rigorous
proof of this composition rule. It is interesting to note that in this general
case the composition rule cannot be reduced to the multiplicative property of
the transfer matrix of the Schr\"{o}dinger equation on the graph.

Such composition rules are important in the study of the electric conduction in
multi-terminal mesoscopic systems. By the Landauer-B\"{u}ttiker theory the electric
conduction in mesoscopic systems is directly related to the transmission
probability and thus to the scattering matrix
\cite{Landauer:70,Buettiker:86,Buettiker:88,Buettiker:90}. A good introduction
into the theory of electronic transport in such systems is given in the book
\cite{Datta} by S.\ Datta. The formal arguments leading to the composition rule
for the scattering matrices are presented on p.\ 125 -- 126 of this book.

The composition rules are also very useful in the study of statistical
properties of large random or periodic systems. Examples of such systems can be
found e.g.\ in \cite{Avishai:Luck,Avron:Exner:Last,Exner:96}. In
\cite{Kostrykin:Schrader:99a,Kostrykin:Schrader:99c,Kostrykin:Schrader:99d,Kostrykin:Schrader:2000}
we proved that in arbitrary dimensions the scattering phase (or more generally
of the spectral shift function) per interaction volume equals (up to a factor
$\pi$) the difference of the integrated densities of states for the free and
interaction theories respectively. In the strictly one-dimensional situation
(Schr\"{o}dinger operators on the line) the Lyapunov exponent is known to be
related to the logarithmic density of transmission probability
\cite{Lifshitz:Gredeskul:Pastur:82,Lifshitz:Gredeskul:Pastur:88,Marchenko:Pastur:86,Kostrykin:Schrader:99a}.
Due to the Ishii-Pastur-Kotani theorem (see e.g.\
\cite{Cycon:Froese:Kirsch:Simon}) the vanishing of the transmission amplitude
for almost all values of energy implies localization (i.e.\ the spectrum must
be purely point), see also the related works
\cite{Tong,Dorokhov:82,Dorokhov:83,Dorokhov:84,Dorokhov:88,Mello:Pereyra:Kumar,Stone:Mello:Muttalib:Pichard,Beenakker}.

Certain Laplace operators on (infinite) periodic graphs were previously
considered in \cite{Avron:Exner:Last,Exner:96}. There are also some attempts to
consider differential operators on regular graphs with random boundary
conditions or on random graphs with deterministic boundary conditions (see
e.g.\ \cite{Avishai:Luck}). Some other examples can be found also in
\cite[Chapter 3]{Habil}. A difference Laplace operator on the edges of
aperiodic tilings was considered in \cite{Kellendonk}. Such systems provide a
main field of application for our composition rule which will be discussed in a
forthcoming publication.

The article is organized as follows. In Section \ref{sec2} we recall the
general quantum scattering theory on graphs as given in
\cite{Kostrykin:Schrader:99b}. In Section \ref{sec3} we recall the definition
of the generalized star product and study some its properties. In particular we
show that this product applies to arbitrary unitary matrices. In Section
\ref{sec4} we give a rigorous proof of the composition rule for scattering
matrices on arbitrary finite graphs. Section \ref{sec5} is devoted to the
special case of graphs having an even number $2p$ of external lines. If the new
graph is obtained by gluing of exactly $p$ lines then it has again $2p$
external lines. We consider the question whether in this case the composition
rule for the scattering matrices can be reduced to the multiplication rule of
the corresponding transfer matrices. In general for $p>1$ the answer is
negative. We formulate a necessary and sufficient condition, which guarantees
that the composition rule for the scattering matrices is equivalent to the
standard multiplication of transfer matrices.

After completing the work we received the preprint \cite{Harmer} by M.\ Harmer
where among other questions the composition rule for the scattering matrices is
also considered. The results there partially recover our Theorem
\ref{graph:thm:factorization} below.

We are indebted to P.\ Kuchment for sending us the preliminary version of the
preprint \cite{Kuchment:Zeng} and also for pointing out the works of R.\
Carlson \cite{Carlson:97,Carlson:98,Carlson:99}.

\section{The Laplacian on a Graph and its Scattering Matrix}\label{sec2}
\setcounter{equation}{0}

In this section we will recall the definition of Schr\"{o}dinger operators on an
arbitrary but finite graph and the construction of their scattering matrices
\cite{Kostrykin:Schrader:99b}.

We consider an arbitrary graph $\Gamma$ with a finite number $n\geq 1$ of
external and a finite number $m\geq 0$ of internal lines (edges). More
precisely this means that outside of a finite domain the graph is isomorphic to
the union of $n$ positive half-lines. Any internal line ends at two, not
necessary different vertices and has a finite length. We assume that any vertex
of $\Gamma$ has non-zero degree, i.e.\ for any vertex there is at least one
edge (internal or external) with which it is incident.

Let the set $\cE$ label the external and the set $\cI$ the internal lines of
the graph. We assume that the sets $\cE$ and $\cI$ are ordered in an arbitrary
but fixed way. To each $e\in\cE$ we associate the infinite interval
$[0,\infty)$ and to each $i\in\cI$ the finite directed interval $[0,a_i]$,
where $a_i>0$ is the length of this line. With this association the graph
becomes directed, such that the initial vertex of an edge of length $a$
corresponds to $x=0$ and the terminal vertex corresponds to $x=a$. The external
lines are assumed to be directed in the positive direction of half-lines.

We define the Hilbert space $\cH=L^2(\Gamma)$ as
\begin{equation*}
\cH=\cH_{\cE}\ \oplus\ \cH_{\cI}, \qquad \cH_{\cE}=\bigoplus_{e\in\cE}\cH_{e},
\qquad \cH_{\cI}=\bigoplus_{i\in\cI}\cH_{i},
\end{equation*}
where $\cH_e=L^2(0,\infty)$ and $\cH_i=L^2(0,a_i)$. Elements of $\cH$ are
written as column vectors
\begin{equation}\label{graph:ext:int}
\psi=\left(\{\psi_e\}_{e\in\cE},\ \{\psi_i\}_{i\in\cI}\right)^T=(\psi_\cE,\psi_\cI)^T,
\qquad \psi_e\in\cH_e,\qquad \psi_i\in\cH_i.
\end{equation}
Similarly we define the Sobolev space $W^{2,2}(\Gamma)$ as
\begin{equation*}
W^{2,2}(\Gamma)=\bigoplus_{e\in\cE}W^{2,2}(0,\infty)\
\oplus\ \bigoplus_{i\in\cI} W^{2,2}(0,a_i),
\end{equation*}
where $W^{2,2}(0,\infty)$ and $W^{2,2}(0, a_i)$ are the usual Sobolev spaces of
square integrable functions whose distributional second derivatives are also
square integrable (see e.g.\ \cite{RS2}). Let $\left[\ \right]:
W^{2,2}(\Gamma)\rightarrow
\C^{2(n+2m)}$ be the surjective linear map which associates to each $\psi$ the
element $[\psi]$ given as
\begin{equation*}
[\psi]=\begin{pmatrix} (\{\psi_e(0)\}_{e\in\cE}, \{\psi_i(0)\}_{i\in\cI},\
\{\psi_i(a_i)\}_{i\in\cI})^T\\
(\{\psi'_e(0)\}_{e\in\cE}, \{\psi'_i(0)\}_{i\in\cI},\
\{-\psi'_i(a_i)\}_{i\in\cI})^T
\end{pmatrix}=\begin{pmatrix}\underline{\psi} \\ \underline{\psi}'
\end{pmatrix}
\end{equation*}
again viewed as a column vector with the same ordering as in $\psi$, i.e.\ with
the ordering given by the ordering of $\cE$ and $\cI$.

In \cite{Kostrykin:Schrader:99b} we showed that for any two
$(n+2m)\times(n+2m)$ complex matrices $A$ and $B$ with $AB^\ast$ being
Hermitian and the $(n+2m)\times 2(n+2m)$ matrix $(A,B)$ having maximal rank
equal to $n+2m$, one can define the self-adjoint Laplace operator
 $\Delta(A,B,\underline{a})$ in $\cH$ corresponding to the
boundary condition
\begin{equation}\label{2}
A\underline{\psi}+B\underline{\psi}^\prime=0.
\end{equation}
Here $\underline{a}=(a_1,\ldots,a_m)^T\in\R_+^m$, $m=\#(\cI)$. Furthermore, any
self-adjoint extension of the Laplacian on the given graph is given by
$\Delta(A,B,\underline{a})$ with some matrices $A$ and $B$ satisfying the
properties stated above. If $\cI=\emptyset$ we simply write $\Delta(A,B)$
instead of $\Delta(A,B,\cdot)$.

Before we turn to the scattering theory for $\Delta(A,B,\underline{a})$ we
recall some well-known facts from scattering theory in two Hilbert spaces
$\fH_1$ and $\fH_2$ (see e.g.\ \cite{RS3}). Let $H_1$ and $H_2$ be self-adjoint
operators in the Hilbert spaces $\fH_1$ and $\fH_2$ respectively. Let $\cJ$ be
a bounded operator from $\fH_1$ into $\fH_2$. The two-space wave operators are
defined as the strong limits
\begin{displaymath}
\Omega^\pm(H_2,H_1;\cJ)= \slim_{t\rightarrow\mp\infty}
e^{iH_2 t} \cJ e^{-i H_1t} P_{\mathrm{ac}}(H_1),
\end{displaymath}
where $P_{\mathrm{ac}}(H)$ denotes the projection onto the absolute continuous
subspace of $H$. We consider the sets $\fN_\pm$ of elements $g\in\fH_2$ for
which $\lim_{t\rightarrow\mp\infty}\|\cJ^\ast e^{-iH_2 t} P_{\mathrm{ac}}(H_2)
g\|=0$. The wave operators $\Omega^\pm$ are called $\cJ$-complete if
$\fH_2=\overline{\Ran(\Omega^\pm)}\oplus\fN_\pm$. For details we refer e.g.\ to
Chapter 3 of the book \cite{Yafaev}.

Now we consider a graph $\Gamma_{\cE}$ consisting of the external lines of the
original graph $\Gamma$. On the graph $\Gamma_{\cE}$ we consider the operator
$-\Delta(A_{\cE}=0, B_{\cE}=\1)$ corresponding to Neumann boundary conditions.
Let $\cJ:\ \cH\rightarrow \cH_{\cE}$ be given as $\cJ\psi=\psi_{\cE}$ according
to the notation \eqref{graph:ext:int}. In particular $\cJ$ is identity if
$m=0$. Since we actually deal with finite dimensional perturbations by the
Kato-Rosenblum theorem (see e.g.\ \cite[Theorem 6.2.3 and Corollary
6.2.4]{Yafaev}) the wave operators $\Omega^\pm(-\Delta(A,B,\underline{a}),
-\Delta(A_{\cE}=0, B_{\cE}=\1);\cJ)$ exist and are $\cJ$-complete. Thus the scattering
operator
\begin{equation}\label{S.matrix:2spaces}
S(-\Delta(A,B,\underline{a}),-\Delta(A_{\cE}=0,B_{\cE}=\1);\cJ)=
(\Omega^-)^\ast\Omega^+:\cH_{\cE}\rightarrow\cH_{\cE}
\end{equation}
is unitary and its layers $S_{A,B,\underline{a}}(\lambda):\
\C^n\rightarrow\C^n$ (in the direct integral representation with respect to
$-\Delta(A_{\cE}=0,B_{\cE}=\1)$) are also unitary for almost all energies
$\lambda\in\R_+$.

The resulting scattering matrix is related to the scattering wave function for
the operator $-\Delta(A,B,\underline{a})$ at energy $\lambda>0$ as follows. The
function $\psi^{k}(\cdot,\lambda)$ indexed by $k\in\cE$ and with components
\begin{equation}\label{10}
\psi^{k}_{j}(x,\lambda)=\begin{cases}S_{jk}(\lambda)e^{i\sqrt{\lambda}x}&\text{for}
                                          \;j\in\cE,j\neq k,\\
  e^{-i\sqrt{\lambda}x}+S_{kk}(\lambda)e^{i\sqrt{\lambda}x}&\text{for}\;j\in\cE,j=k\\
                                  \alpha_{jk}(\lambda)e^{i\sqrt{\lambda}x}+
        \beta_{jk}(\lambda)e^{-i\sqrt{\lambda}x}&\text{for}\;j\in\cI,\end{cases},
\end{equation}
solves the Schr\"{o}dinger equation with the operator $-\Delta(A,B,\underline{a})$
for energy $\lambda>0$.

Recall that in potential scattering for one particle Schr\"{o}dinger operators the
wave operators give precise meaning to the scattering wave functions, i.e.\
solutions of the Schr\"{o}dinger equation at positive energy $\lambda>0$. Similarly
the wave operators $\Omega^\pm(-\Delta(A,B,\underline{a}),
-\Delta(A_{\cE}=0, B_{\cE}=\1);\cJ)$ determine the ``external part" of the
scattering wave function, i.e.\ $\psi_j^k(x,\lambda)$ for $j\in\cE$. The
completness of usual wave operators means that any solution of the Schr\"{o}dinger
equation at energy $\lambda>0$ can be uniquely represented as a superposition
of the scattering wave functions. Similarly, in the present context the
$\cJ$-completness of the wave operators $\Omega^\pm(-\Delta(A,B,\underline{a}),
-\Delta(A_{\cE}=0, B_{\cE}=\1);\cJ)$ means that the external part of any solution
for the Schr\"{o}dinger equation with the operator $-\Delta(A,B,\underline{a})$ at
energy $\lambda>0$ can uniquely be represented as a linear combination of the
external parts of the scattering wave functions \eqref{10}.

The scattering matrix $S_{A,B,\underline{a}}(\lambda)$ as well as the $m\times
n$ matrix amplitudes $\alpha_{A,B,\underline{a}}(\lambda)$ and
$\beta_{A,B,\underline{a}}(\lambda)$ are determined as solutions to the
equation
\begin{equation}\label{11}
Z_{A,B,\underline{a}}(\lambda)\begin{pmatrix} S(\lambda)\\
                      \alpha(\lambda)\\
                    \beta(\lambda)\end{pmatrix} =-(A-i\sqrt{\lambda}B)
                    \begin{pmatrix} \1\\
                               0\\
                               0 \end{pmatrix}
\end{equation}
where
\begin{equation}\label{Z.def}
  Z_{A,B,\underline{a}}(\lambda)=AX_{\underline{a}}(\lambda)
  +i\sqrt{\lambda}BY_{\underline{a}}(\lambda)
\end{equation}
is the $(n+2m)\times (n+2m)$ matrix with
\begin{displaymath}
 X_{\underline{a}}(\lambda)= \begin{pmatrix}\1&0&0\\
                                  0&\1&\1\\
               0&e^{i\sqrt{\lambda}\underline{a}}&e^{-i\sqrt{\lambda}\underline{a}}
               \end{pmatrix}
\end{displaymath}
and
\begin{displaymath}
 Y_{\underline{a}}(\lambda)= \begin{pmatrix}\1&0&0\\
                                  0&\1&-\1\\
               0&-e^{i\sqrt{\lambda}\underline{a}}&e^{-i\sqrt{\lambda}\underline{a}}
               \end{pmatrix}.
\end{displaymath}
Here $e^{\pm i\sqrt{\lambda}\underline{a}}$ stands for the $m\times m$ diagonal
matrix with elements
\begin{equation*}
\left(e^{\pm i\sqrt{\lambda}\underline{a}}\right)_{jk}=e^{\pm i\sqrt{\lambda}a_j}\delta_{jk},
\qquad j,k\in\cI.
\end{equation*}

If  $\det Z_{A,B,\underline{a}}(\lambda)\neq 0$ the scattering matrix
$S(\lambda)=S_{A,B,\underline{a}}(\lambda)$ as well as the $m\times n$ matrices
$\alpha(\lambda)=\alpha_{A,B,\underline{a}}(\lambda)$ and
$\beta(\lambda)=\beta_{A,B,\underline{a}}(\lambda)$ can be uniquely determined
by solving the equation
\eqref{11} in the form
\begin{equation}\label{cont}
\begin{pmatrix} S(\lambda) \\ \alpha(\lambda) \\ \beta(\lambda) \end{pmatrix}=
-Z_{A,B,\underline{a}}(\lambda)^{-1}(A-i\sqrt{\lambda}B)
\begin{pmatrix} \1 \\ 0 \\ 0\end{pmatrix}.
\end{equation}
We denote by $\Sigma_{A,B,\underline{a}}=\left\{\lambda>0\ |\ \det
Z_{A,B,\underline{a}}(\lambda)=0
\right\}$ the set of exceptional points for which $Z_{A,B,\underline{a}}(\lambda)$ is not
invertible.

Let $\phi$ be an arbitrary measurable function on the graph $\Gamma$. The
subset $\supp\ \phi$ of all edges of the graph $\Gamma$ will be called the
\emph{support} of the function $\phi$ if $\phi\neq 0$ a.e.\ on $\supp\ \phi$.

In \cite{Kostrykin:Schrader:99b} we proved the following

\begin{theorem}\label{graph:theorem:2}
For any boundary condition $(A,B)$ and arbitrary $\underline{a}\in\R_+^m$ the
set $\Sigma_{A,B,\underline{a}}$ equals the set $\sigma_{A,B,\underline{a}}$ of
all positive eigenvalues of $-\Delta(A,B,\underline{a})$. This set is discrete
and has no finite accumulation points in $\R_+$. The eigenfunctions of
$-\Delta(A,B,\underline{a})$ have support on the internal lines of the graph.
For all $\lambda\in\Sigma_{A,B,\underline{a}}$ the equation \eqref{11} is still
solvable and determines $S_{A,B,\underline{a}}(\lambda)$ uniquely.
\end{theorem}

Given an arbitrary $n\times n$ unitary matrix $U$ and an energy $\lambda_0>0$
we can find boundary conditions $A,B$ defining a self-adjoint operator
$-\Delta(A,B)$ on a single-vertex graph (i.e.\ with $m=0$) with $n$ external
lines such that the corresponding scattering matrix is given as
$S_{A,B}(\lambda_0)=U$. The proof of this fact can be found in
\cite{Kostrykin:Schrader:99e}. For other inverse problems on graphs we refer to
\cite{Gerasimenko,Carlson:99}.

Recall that by definition the operator $\Delta(A,B,\underline{a})$ is real if
it commutes with complex conjugation, i.e.\ for any
$\psi\in\cD(\Delta(A,B,\underline{a}))$ the function $\overline{\psi}$ also
belongs to $\cD(\Delta(A,B,\underline{a}))$ and
$\Delta(A,B,\underline{a})\psi=\overline{\Delta(A,B)\overline{\psi}}$.
Equivalently, this means that any $\psi\in\cD(\Delta(A,B,\underline{a}))$
(i.e.\ $\psi\in W^{2,2}(\Gamma)$ satisfying
$A\underline{\psi}+B\underline{\psi}^\prime=0$) also satisfies the equation
$\overline{A}\underline{\psi}+\overline{B}\underline{\psi}^\prime=0$. Thus,
$\Delta(A,B,\underline{a})$ is real iff $\Ker (A,B)=\Ker
(\overline{A},\overline{B})$. The last condition is satisfied iff there is an
invertible matrix $C_1$ such that $A=C_1 \overline{A}$, $B=C_1 \overline{B}$ or
alternatively there is an invertible matrix $C_2$ such that both $C_2 A$ and
$C_2 B$ are real. We recall that
$\Delta(A,B,\underline{a})=\Delta(CA,CB,\underline{a})$ for any invertible $C$
(see \cite{Kostrykin:Schrader:99b}).

In \cite[Corollary 3.2]{Kostrykin:Schrader:99b} we have proved the following

\begin{theorem}\label{transp}
For arbitrary $\underline{a}\in\R_+^m$, $\lambda>0$, and all boundary
conditions $A,B$ defining the self-adjoint operator $\Delta(A,B,\underline{a})$
\begin{equation}\label{sym.S}
S_{\overline{A},\overline{B},\underline{a}}(\lambda)^T =
S_{A,B,\underline{a}}(\lambda).
\end{equation}
In particular, if the operator $\Delta(A,B,\underline{a})$ is real then
$S_{A,B,\underline{a}}(\lambda)=S_{A,B,\underline{a}}(\lambda)^T$ for all
$\lambda>0$.
\end{theorem}

Here we give an alternative proof.

\begin{proof}
From the selfadjointness of the operator $\Delta(A,B,\underline{a})$ it follows
that the matrix $A+i\sqrt{\lambda}B$ is invertible for all $\lambda>0$ (see
\cite{Kostrykin:Schrader:99b}). The relation
\eqref{11} implies that
\begin{equation*}
\begin{pmatrix}S_{A,B,\underline{a}}(\lambda) \\ \alpha_{A,B,\underline{a}}(\lambda) \\
e^{-i\sqrt{\lambda}\underline{a}}\beta_{A,B,\underline{a}}(\lambda)\end{pmatrix}=
-(A+i\sqrt{\lambda}B)^{-1}(A-i\sqrt{\lambda}B)\begin{pmatrix}\1 \\
\beta_{A,B,\underline{a}}(\lambda) \\ e^{i\sqrt{\lambda}\underline{a}}
\alpha_{A,B,\underline{a}}(\lambda)
\end{pmatrix}.
\end{equation*}
Similarly, for the operator $\Delta(\overline{A},\overline{B},\underline{a})$
we have
\begin{equation*}
\begin{pmatrix}S_{\overline{A},\overline{B},\underline{a}}(\lambda) \\
\alpha_{\overline{A},\overline{B},\underline{a}}(\lambda) \\
e^{-i\sqrt{\lambda}\underline{a}}\beta_{\overline{A},\overline{B},\underline{a}}(\lambda)
\end{pmatrix}=
-(\overline{A}+i\sqrt{\lambda}\overline{B})^{-1}
(\overline{A}-i\sqrt{\lambda}\overline{B})\begin{pmatrix}\1 \\
\beta_{\overline{A},\overline{B},\underline{a}}(\lambda) \\
e^{i\sqrt{\lambda}\underline{a}}\alpha_{\overline{A},\overline{B},\underline{a}}(\lambda)
\end{pmatrix}.
\end{equation*}
Taking the complex conjugate and multiplying by
$(A+i\sqrt{\lambda}B)^{-1}(A-i\sqrt{\lambda}B)$ from the left we obtain
\begin{equation*}
\begin{pmatrix}\1 \\[1mm]
\overline{\beta_{\overline{A},\overline{B},\underline{a}}(\lambda)} \\[1mm]
e^{-i\sqrt{\lambda}\underline{a}}
\overline{\alpha_{\overline{A},\overline{B},\underline{a}}(\lambda)}\end{pmatrix}=
-(A+i\sqrt{\lambda}B)^{-1}
(A-i\sqrt{\lambda}B)\begin{pmatrix}
\overline{S_{\overline{A},\overline{B},\underline{a}}(\lambda)}
\\[1mm]
\overline{\alpha_{\overline{A},\overline{B},\underline{a}}(\lambda)} \\[1mm]
e^{i\sqrt{\lambda}\underline{a}}
\overline{\beta_{\overline{A},\overline{B},\underline{a}}(\lambda)}
\end{pmatrix}.
\end{equation*}
We multiply this relation by $S_{\overline{A},\overline{B}}(\lambda)^T$ from
the right and make use of the unitarity of the scattering matrix in the form
$\overline{S_{\overline{A},\overline{B}}(\lambda)}S_{\overline{A},\overline{B}}(\lambda)^T=\1$
thus obtaining
\begin{equation*}
\begin{pmatrix}S_{\overline{A},\overline{B},\underline{a}}(\lambda)^T \\[1mm]
\overline{\beta_{\overline{A},\overline{B},\underline{a}}(\lambda)}
S_{\overline{A},\overline{B},\underline{a}}(\lambda)^T \\[1mm]
e^{-i\sqrt{\lambda}\underline{a}}
\overline{\alpha_{\overline{A},\overline{B},\underline{a}}(\lambda)}
S_{\overline{A},\overline{B},\underline{a}}(\lambda)^T\end{pmatrix}=
-(A+i\sqrt{\lambda}B)^{-1}
(A-i\sqrt{\lambda}B)\begin{pmatrix}
\1 \\[1mm]
\overline{\alpha_{\overline{A},\overline{B},\underline{a}}(\lambda)}
S_{\overline{A},\overline{B},\underline{a}}(\lambda)^T \\[1mm]
e^{i\sqrt{\lambda}\underline{a}}\overline{\beta_{\overline{A},\overline{B},\underline{a}}(\lambda)}
S_{\overline{A},\overline{B},\underline{a}}(\lambda)^T
\end{pmatrix}.
\end{equation*}
Equivalently, this relation can be written in a form analogous to \eqref{11},
\begin{equation}\label{ref.S}
Z_{A,B,\underline{a}}(\lambda)\begin{pmatrix}
S_{\overline{A},\overline{B},\underline{a}}(\lambda)^T
\\[1mm]
                      \overline{\beta_{\overline{A},\overline{B},\underline{a}}(\lambda)}
                      S_{\overline{A},\overline{B},\underline{a}}(\lambda)^T\\[1mm]
                    \overline{\alpha_{\overline{A},\overline{B},\underline{a}}(\lambda)}
                    S_{\overline{A},\overline{B},\underline{a}}(\lambda)^T\end{pmatrix}
                    =-(A-i\sqrt{\lambda}B)
             \begin{pmatrix} \1\\
                               0\\
                               0 \end{pmatrix}.
\end{equation}
In \cite{Kostrykin:Schrader:99b} we proved that the equation \eqref{11} has a
solution for all $\lambda>0$. If $\lambda>0$ is not an eigenvalue of the
operator $\Delta(A,B,\underline{a})$ then it has a unique solution. If
$\lambda>0$ is an eigenvalue of $\Delta(A,B,\underline{a})$ this solution is
non-unique, but still determines the scattering matrix uniquely. Therefore from
comparison of \eqref{11} and \eqref{ref.S} the relation \eqref{sym.S} follows.
If $\Delta(A,B,\underline{a})$ is real by the remark preceding the theorem we
can choose the matrices $A$ and $B$ to be real. From this and from
\eqref{sym.S} the second claim of the theorem follows.
\end{proof}

We note that the comparison of \eqref{11} with \eqref{ref.S} also gives the
relations
\begin{eqnarray*}
\alpha_{A,B,\underline{a}}(\lambda) &=&
\overline{\beta_{\overline{A},\overline{B},\underline{a}}(\lambda)}\
S_{A,B,\underline{a}}(\lambda)^T,\\
\beta_{A,B,\underline{a}}(\lambda) &=&
\overline{\alpha_{\overline{A},\overline{B},\underline{a}}(\lambda)}\
S_{A,B,\underline{a}}(\lambda)^T.
\end{eqnarray*}

\section{The Generalized Star Product}\label{sec3}
\setcounter{equation}{0}

Let $U^{(1)}$ and $U^{(2)}$ be arbitrary $n_1\times n_1$ and $n_2\times n_2$
unitary matrices respectively. Let $p$ be some integer satisfying $1\leq
p<(n_1+n_2)/2$, $p\leq n_j$, $j=1,2$, and $V$ be an arbitrary $p\times p$
unitary matrix. We write $U^{(1)}$ and $U^{(2)}$ in a $2\times 2$-block form
\begin{equation}\label{54}
U^{(1)}= \begin{pmatrix} U^{(1)}_{11}&U^{(1)}_{12}\\
                        U^{(1)}_{21}&U^{(1)}_{22}\end{pmatrix},\qquad
U^{(2)}
= \begin{pmatrix} U^{(2)}_{11}&U^{(2)}_{12}\\
       U^{(2)}_{21}&U^{(2)}_{22}\end{pmatrix},
\end{equation}
where $U^{(1)}_{22}$ and $U^{(2)}_{11}$ are $p\times p$ matrices,
$U^{(1)}_{11}$ is an $(n_1-p)\times (n_1-p)$ matrix, $U^{(2)}_{22}$ is an
$(n_2-p)\times (n_2-p)$ matrix etc. The unitarity condition for $U^{(1)}$ then
reads
\begin{eqnarray*} U^{{(1)}^{\ast}}_{11}U^{(1)}_{11}+
  U^{{(1)}^{\ast}}_{21}U^{(1)}_{21}&=&\1,\\
U^{{(1)}^{\ast}}_{12}U^{(1)}_{12}+
  U^{{(1)}^{\ast}}_{22}U^{(1)}_{22}&=&\1,\\
U^{{(1)}^{\ast}}_{11}U^{(1)}_{12}+
  U^{{(1)}^{\ast}}_{21}U^{(1)}_{22}&=&0,\\
U^{{(1)}^{\ast}}_{12}U^{(1)}_{11}+
  U^{{(1)}^{\ast}}_{22}U^{(1)}_{21}&=&0
\end{eqnarray*}
and similarly for $U^{(2)}$.

\begin{definition}
The unitary matrix $U^{(1)}$ is called \textbf{$V$-compatible} with the unitary
matrix $U^{(2)}$ if the $p\times p$ matrix $VU^{(1)}_{22}V^\ast U_{11}^{(2)}$
does not have $1$ as an eigenvalue. For the case $V=\1$\hspace{2mm} the matrix
$U^{(1)}$ is simply called
\textbf{compatible} with $U^{(2)}$ (for given $p\geq 1$).
\end{definition}

Note that the compatibility of the matrices is not a symmetric relation, i.e.\
if $U^{(1)}$ is $V$-compatible with $U^{(2)}$ then $U^{(2)}$ need not be
$V$-compatible with $U^{(1)}$.

Obviously, if $U^{(1)}$ is $V$-compatible with $U^{(2)}$ then also the matrix
$V^\ast U_{11}^{(2)}VU^{(1)}_{22}$ does not have $1$ as an eigenvalue. Indeed,
let us assume the converse, i.e.\ let there be non-zero $c\in\C^p$ such that
$V^\ast U_{11}^{(2)}V U_{22}^{(1)}c=c$ and thus
\begin{equation*}
V U_{22}^{(1)} V^\ast U_{11}^{(2)}V U_{22}^{(1)}c=V U_{22}^{(1)}c.
\end{equation*}
Since $V U_{22}^{(1)}c\neq 0$ the matrix $V U_{22}^{(1)} V^\ast U_{11}^{(2)}$
has $1$ as an eigenvalue, which is a contradiction. From this it also follows
that if $U^{(1)}$ is not $V$-compatible with $U^{(2)}$ then the matrix $V^\ast
U_{11}^{(2)} V U_{22}^{(1)}$ has 1 as an eigenvalue.

From the unitarity conditions it follows that
\begin{eqnarray*}
0\leq U^{{(1)}^\ast}_{11}U^{(1)}_{11} = \1-U^{{(1)}^\ast}_{21}U^{(1)}_{21}\leq
\1,\\
0\leq U^{{(1)}^\ast}_{22}U^{(1)}_{22} = \1-U^{{(1)}^\ast}_{12}U^{(1)}_{12}\leq
\1,
\end{eqnarray*}
and thus $\|U^{(1)}_{11}\|\leq 1$, $\|U^{(1)}_{22}\|\leq 1$. Similar
inequalities hold for $U^{(2)}_{11}$ and $U^{(2)}_{22}$. Therefore
$\|VU^{(1)}_{22}V^\ast U_{11}^{(2)}\|\leq 1$. Strict inequality holds whenever
$\| U^{(1)}_{22}\| <1$ or $\| U_{11}^{(2)}\| <1$ and then $U^{(1)}$ is
$V$-compatible with $U^{(2)}$ and $U^{(2)}$ is $V$-compatible with $U^{(1)}$
for all unitary $p\times p$ matrices  $V$.

In general if $U^{(1)}$ is $V$-compatible with $U^{(2)}$ then it is easy to see
that the following $p\times p$ matrices exist:
\begin{equation}\label{K.def}
\begin{split}
K_{1}&=\ (\1-VU^{(1)}_{22}V^{\ast}U^{(2)}_{11})^{-1}V
=V(\1-U^{(1)}_{22}V^{\ast}U^{(2)}_{11}V)^{-1},\\
K_{2}&=\ (\1-V^{\ast}U^{(2)}_{11}VU^{(1)}_{22})^{-1}V^{\ast}
=V^{\ast}(\1-U^{(2)}_{11}VU^{(1)}_{22}V^{\ast})^{-1}.
\end{split}
\end{equation}
An easy calculation establishes the following relations
\begin{equation}\label{alt.31}
\begin{split}
K_{1}&=V+VU^{(1)}_{22}V^{\ast}U^{(2)}_{11}K_{1}=
V+VU^{(1)}_{22}K_{2}U^{(2)}_{11}V\\
&=V+K_{1}U^{(1)}_{22}V^{\ast}U^{(2)}_{11}V,\\
K_{2}&=V^{\ast}+V^{\ast}U^{(2)}_{11}VU^{(1)}_{22}K_{2}=
V^{\ast}+V^{\ast}U^{(2)}_{11}K_{1}U^{(1)}_{22}V^{\ast}\\
      &=V^{\ast}+K_{2}U^{(2)}_{11}VU^{(1)}_{22}V^{\ast}.\\
      \end{split}
\end{equation}
Note that formally one has the power series expansion
\begin{eqnarray*}
K_{1}&=&\sum^{\infty}_{m=0}
   (VU^{(1)}_{22}V^{\ast}U^{(2)}_{11})^{m}V,\\
K_{2}&=&\sum^{\infty}_{m=0}
   (V^{\ast}U^{(2)}_{11}VU^{(1)}_{22})^{m}V^{\ast},
\end{eqnarray*}
which is rigorous if $\|U_{22}^{(1)}\|<1$ or $\|U_{11}^{(2)}\|<1$. These power
series expansions combined with the superposition principle were used in
\cite{Kostrykin:Schrader:99b} to give a formal proof that the composition rule
for scattering matrices was given by the generalized star product.

With these preparations the generalized star product $U=U^{(1)}*_V U^{(2)}$ of
the unitary matrices $U^{(1)}$ and $U^{(2)}$ is defined as follows. Write the
$(n_1+n_2-2p)\times(n_1+n_2-2p)$ matrix $U$ in a $2\times 2$ block form as
\begin{displaymath}
        U=\begin{pmatrix}U_{11}&U_{12}\\
                                  U_{21}&U_{22} \end{pmatrix},
\end{displaymath}
where $U_{11}$ is an $(n_1-p)\times (n_1-p)$ matrix, $U_{22}$ is an
$(n_2-p)\times (n_2-p)$ matrix etc. These matrices are now defined as
\begin{equation}\label{56}
\begin{split}
 U_{11}&=U^{(1)}_{11}
        +U^{(1)}_{12}K_{2}U^{(2)}_{11}VU^{(1)}_{21},\\
 U_{22}&=U^{(2)}_{22}
      +U^{(2)}_{21}K_{1}U^{(1)}_{22}V^{\ast}U^{(2)}_{12},\\
 U_{12}&=U^{(1)}_{12}K_{2}U^{(2)}_{12},\\
 U_{21}&=U^{(2)}_{21}K_{1}U^{(1)}_{21}.\\
 \end{split}
\end{equation}
In particular for an arbitrary $n\times n$ unitary matrix $U$ and all $p$ such
that $1\leq p<n$ the $2p\times 2p$ matrices $\E=\begin{pmatrix} 0 & \1 \\ \1 &
0 \end{pmatrix}$ serve as units for the $\ast_V$-product when $V=\1$,
\begin{displaymath}
\begin{pmatrix} 0 & \1 \\ \1 & 0 \end{pmatrix} *_V U =
U *_V \begin{pmatrix} 0 & \1 \\ \1 & 0 \end{pmatrix} = U.
\end{displaymath}

Further we will need the following Perron-Frobenius-type result which for the
sake of generality will be formulated to cover also the infinite-dimensional
case:

\begin{lemma}\label{Frobenius}
Let a compact operator $A$ on a separable Hilbert space $\fH$ be a contraction,
i.e.\ $\|A\|\leq 1$. Suppose that $\lambda=1$ is an eigenvalue of $A$. Then

(i) every $c\in\fH$ such that $Ac=c$ also satisfies $A^\ast c=c$ and hence also
$A^\ast A c=A A^\ast c =c$,

(ii) the geometric and algebraic multiplicities of the eigenvalue $\lambda=1$
are equal.
\end{lemma}

\begin{proof}
The claim (i) is an easy consequence of the singular values decomposition (see
e.g.\ \cite[p.\ 155]{Horn:Johnson}). Thus we have $\Ker (A-1)=\Ker (A^\ast-1)=
\left(\Ran(A-1)\right)^\perp$. The claim (ii) now follows from the fact that
the geometric and algebraic multiplicities of an eigenvalue $\lambda$ are
unequal iff $\Ran(A-\lambda)\cap\Ker(A-\lambda)$ is non-trivial.
\end{proof}

Also we will make use of the following

\begin{lemma}\label{graph:lem:factor}
Let $A$ and $B$ be linear compact operators on a separable Hilbert space $\fH$
such that $\|A\|\leq 1$, $\|B\|\leq 1$ and $ABb=b$ for some $b\in\fH$. Then
$B^\ast Bb=b$.
\end{lemma}

\begin{proof}
Without loss of generality we may assume that $\|b\|_{\fH}=1$. By Lemma
\ref{Frobenius}
\begin{equation}\label{Gleichung:1}
B^\ast A^\ast AB b=b.
\end{equation}
Therefore by well-known inequalities for the singular values of compact
operators (see e.g.\ \cite{Gohberg:Krein}) we have
\begin{equation*}
1\leq s(AB)\leq s(A) s(B)\leq \|A\|\ \|B\|\leq 1,
\end{equation*}
where $s(K)$ denotes the maximal singular value of a compact operator $K$,
i.e.\ the maximal eigenvalue of the self-adjoint non-negative operator $K^\ast
K$. This gives $s(AB)=s(A)=s(B)=1$. From $s(A)=s(B)=1$ it follows that
$\lambda=1$ is a maximal eigenvalue of $A^\ast A$ and $B^\ast B$. By the
min-max principle (see e.g.\ \cite[Theorem XIII.2]{RS4}) any $c\in\fH$,
$\|c\|_{\fH}=1$ maximizing $(c, K^\ast K c)\leq 1$ satisfies $K^\ast K c=c$.
Moreover, if $(c, K^\ast K c)=1$ with some $\|c\|_{\fH}\leq 1$, then $K^\ast K
c=c$ and $\|c\|_{\fH}=1$. Therefore since
\begin{displaymath}
(Bb, A^\ast ABb)=(b, B^\ast A^\ast AB b)=1
\end{displaymath}
we obtain
\begin{equation}\label{Gleichung:2}
A^\ast ABb=Bb
\end{equation}
and $\|Bb\|=1$. The relations \eqref{Gleichung:1} and \eqref{Gleichung:2} imply
that $B^\ast Bb=B^\ast A^\ast ABb=b$.
\end{proof}

Suppose now that the unitary matrix $U^{(1)}$ is not $V$-compatible with
$U^{(2)}$. In this case the linear subspaces of $\C^p$
\begin{equation*}
\begin{split}
\cC= \Ker (\1- V U_{22}^{(1)} V^\ast U_{11}^{(2)}),& \qquad
\widetilde{\cC} = \Ker (\1- U_{22}^{(1)} V^\ast U_{11}^{(2)} V),\\
\cB = \Ker (\1- V^\ast U_{11}^{(2)} V U_{22}^{(1)}),& \qquad
\widetilde{\cB} = \Ker (\1- U_{11}^{(2)} V U_{22}^{(1)} V^\ast)
\end{split}
\end{equation*}
are nontrivial. By Lemma \ref{Frobenius} we also have
\begin{equation*}
\begin{split}
\cC= \Ker (\1- {U_{11}^{(2)}}^\ast V {U_{22}^{(1)}}^\ast V^\ast),& \qquad
\widetilde{\cC} = \Ker (\1- V^\ast {U_{11}^{(2)}}^\ast V {U_{22}^{(1)}}^\ast),\\
\cB = \Ker (\1- {U_{22}^{(1)}}^\ast V^\ast {U_{11}^{(2)}}^\ast V),& \qquad
\widetilde{\cB} = \Ker (\1- V {U_{22}^{(1)}}^\ast V^\ast {U_{11}^{(2)}}^\ast).
\end{split}
\end{equation*}
Obviously $\widetilde{\cC} = V^\ast \cC$ and $\widetilde{\cB} = V\cB$. Since
$V$ is unitary this implies $\dim \widetilde{\cC} = \dim \cC$ and $\dim
\widetilde{\cB} = \dim \cB$. Furthermore we have

\begin{lemma}\label{graph:neu:i:v}
The subspaces $\cB$ and $\cC$ have equal dimensions, $\dim \cB = \dim \cC$.
Moreover

\begin{equation*}
\begin{array}{lcc}
(i) & \begin{array}{ccccc}
\cB & = & \linspan\{V^\ast U_{11}^{(2)}c,\quad c\in\cC \} & =
   & \linspan\{{U_{22}^{(1)}}^\ast V^\ast c,\quad c\in\cC \},\\
\cC & = & \linspan\{V U_{22}^{(1)}b,\quad b\in\cB \} & =
   & \linspan\{{U_{11}^{(2)}}^\ast V b,\quad b\in\cB \},
\end{array} & \qquad\qquad
\end{array}
\end{equation*}
and
\begin{equation*}
\begin{array}{lcc}
\ \ \ & \begin{array}{ccccc}
\widetilde{\cB} & = & \linspan\{U_{11}^{(2)} V \widetilde{c},\quad \widetilde{c}\in\widetilde{\cC} \} & =
   & \linspan\{V {U_{22}^{(1)}}^\ast \widetilde{c},\quad \widetilde{c}\in\widetilde{\cC} \},\\
\widetilde{\cC} & = & \linspan\{U_{22}^{(1)}V^\ast \widetilde{b},\quad \widetilde{b}\in\widetilde{\cB} \} & =
   & \linspan\{V^\ast {U_{11}^{(2)}}^\ast \widetilde{b},\quad \widetilde{b}\in\widetilde{\cB} \},
\end{array} & \qquad\qquad
\end{array}
\end{equation*}

(ii) $U_{21}^{(2)}c=0$ for all $c\in\cC$,

(iii) $U_{12}^{(1)}b=0$ for all $b\in\cB$,

(iv) ${U_{21}^{(1)}}^\ast \widetilde{c}=0$ for all
$\widetilde{c}\in\widetilde{\cC}$,

(v) ${U_{12}^{(2)}}^\ast \widetilde{b}=0$ for all
$\widetilde{b}\in\widetilde{\cB}$.
\end{lemma}

\begin{proof}
Let $c_i\in\C^p$, $i=1,\ldots,k\leq p\ (k\geq 1)$ be a (not necessarily
orthogonal) basis in $\cC$. For all $i=1,\ldots,k$ we have
\begin{equation}\label{u221}
(\1-V U_{22}^{(1)}V^\ast U_{11}^{(2)})c_i=0.
\end{equation}
Multiplying these equations by $V^\ast U_{11}^{(2)}$ from the left we obtain
\begin{equation*}
(\1 - V^\ast U_{11}^{(2)} V U_{22}^{(1)}) V^\ast U_{11}^{(2)} c_i =0.
\end{equation*}
Thus
\begin{equation}\label{Neu.Incl.1}
\linspan\{V^\ast U_{11}^{(2)}c,\quad c\in\cC \} \subseteq \cB.
\end{equation}
By Lemma \ref{graph:lem:factor} it follows from \eqref{u221} that
\begin{equation}\label{u112}
{U_{11}^{(2)}}^\ast U_{11}^{(2)}c_i=c_i,\qquad i=1,\ldots,k.
\end{equation}
Hence
\begin{equation}\label{Neu.Equal.1}
\dim\linspan\{V^\ast U_{11}^{(2)}c,\quad c\in\cC \}=\dim\cC
\end{equation}
and therefore by \eqref{Neu.Incl.1} $\dim\cC\leq\dim\cB$.

Let $b_i\in\C^p$, $i=1,\ldots,k^\prime\leq p$ be some basis in $\cB$. We have
\begin{equation}\label{3.6.Extra}
(\1 - V^\ast U_{11}^{(2)}V U_{22}^{(1)})b_i=0
\end{equation}
for all $i=1,\ldots,k^\prime$. Multiplying these equations by $V U_{22}^{(1)}$
we obtain
\begin{equation*}
(\1 - V U_{22}^{(1)} V^\ast U_{11}^{(2)}) V U_{22}^{(1)} b_i=0,
\end{equation*}
and thus
\begin{equation}\label{Neu.Incl.2}
\linspan\{V U_{22}^{(1)}b,\quad b\in\cB \}\subseteq \cC.
\end{equation}
Again by Lemma \ref{graph:lem:factor} it follows from \eqref{3.6.Extra} that
\begin{equation}\label{3.7.Extra}
{U_{22}^{(1)}}^\ast U_{22}^{(1)} b_i =b_i,\qquad i=1,\ldots,k^\prime.
\end{equation}
Thus
\begin{equation}\label{Neu.Equal.2}
\dim\linspan\{V U_{22}^{(1)}b,\quad b\in\cB \}=\dim\cB
\end{equation}
and therefore by \eqref{Neu.Incl.2} $\dim\cB\leq\dim\cC$. So we have proved
that $\dim\cB = \dim\cC$. The inclusion \eqref{Neu.Incl.1} and the equality
\eqref{Neu.Equal.1} imply that
\begin{equation*}
\linspan\{V^\ast U_{11}^{(2)}c,\quad c\in\cC \} = \cB.
\end{equation*}
The inclusion \eqref{Neu.Incl.2} and the equality \eqref{Neu.Equal.2} imply
that
\begin{equation*}
\linspan\{V U_{22}^{(1)}b,\quad b\in\cB\} =\cC.
\end{equation*}
The proof of the relations
\begin{eqnarray*}
\cB &=& \linspan\{{U_{22}^{(1)}}^\ast V^\ast c,\quad c\in\cC \},\\
\cC &=& \linspan\{{U_{11}^{(2)}}^\ast V b,\quad b\in\cB \}
\end{eqnarray*}
is similar and will therefore be omitted.

We turn to the proof of (ii) -- (v). By the unitarity of $U^{(2)}$ from
\eqref{u112} it follows that ${U_{21}^{(2)}}^\ast U_{21}^{(2)}c_i=0$.
Since $\Ker A^\ast A=\Ker A$ for any linear operator $A$ we obtain the claim
(ii). By the unitarity of $U^{(1)}$ from \eqref{3.7.Extra} it follows that
${U_{12}^{(1)}}^\ast U_{12}^{(1)} b_i=0$ which proves the claim (iii).

As already noted the vectors $c_i$ and $b_i$ also satisfy
\begin{equation*}
(\1-{U_{11}^{(2)}}^\ast V {U_{22}^{(1)}}^\ast V^\ast) c_i=0,\qquad
(\1-{U_{22}^{(1)}}^\ast V^\ast {U_{11}^{(2)}}^\ast V) b_i=0.
\end{equation*}
A final application of Lemma \ref{graph:lem:factor}  yields
\begin{equation*}
V U_{22}^{(1)} {U_{22}^{(1)}}^\ast V^\ast c_i=c_i,\qquad V^\ast U_{11}^{(2)}
{U_{11}^{(2)}}^\ast V b_i=b_i
\end{equation*}
which by the unitarity of $U^{(1)}, U^{(2)}$ and $V$ implies (iv) and (v) which
completes the proof of the lemma.
\end{proof}

\begin{lemma}\label{X.X.neu}
(i) The matrices $\1-V U_{22}^{(1)} V^\ast U_{11}^{(2)}$ and
$\1-{U_{11}^{(2)}}^\ast V {U_{22}^{(1)}}^\ast V^\ast$ map $\cC^\perp$
bijectively onto itself,

(ii) the matrices $\1- V^\ast U_{11}^{(2)} V U_{22}^{(1)}$ and $\1-
{U_{22}^{(1)}}^\ast V^\ast {U_{11}^{(2)}}^\ast V$ map $\cB^\perp$ bijectively
onto itself,

(iii) the matrices $\1-U_{22}^{(1)} V^\ast U_{11}^{(2)} V$ and $\1-V^\ast
{U_{11}^{(2)}}^\ast V {U_{22}^{(1)}}^\ast$ map $\widetilde{\cC}^\perp$
bijectively onto itself,

(iv) the matrices $\1- U_{11}^{(2)} V U_{22}^{(1)} V^\ast$ and $\1- V
{U_{22}^{(1)}}^\ast V^\ast {U_{11}^{(2)}}^\ast$ map $\widetilde{\cB}^\perp$
bijectively onto itself.
\end{lemma}

\begin{proof}
Since $V$ is unitary by the definitions of $\widetilde{\cB}$ and
$\widetilde{\cC}$ it suffices to prove (i) and (ii). By the definition of $\cC$
we have that
\begin{equation*}
(c, V U_{22}^{(1)} V^\ast U_{11}^{(2)} c_\perp) = ({U_{11}^{(2)}}^\ast V
{U_{22}^{(1)}}^\ast V^\ast c, c_\perp) = (c, c_\perp) =0
\end{equation*}
for any $c\in\cC$ and any $c_\perp\in\cC^\perp$. Thus $(\1-V U_{22}^{(1)}
V^\ast U_{11}^{(2)}) c_\perp \in\cC^\perp$ for all $c_\perp\in\cC^\perp$.
Conversely, by Lemma \ref{Frobenius} for any $c_\perp\in\cC^\perp$ there is a
unique $d\in\cC^\perp$ which satisfies the equation $(\1-V U_{22}^{(1)} V^\ast
U_{11}^{(2)})d= c_\perp$. This proves the claim (i). The claim (ii) is proved
similarly.
\end{proof}

\begin{theorem}\label{graph:dosta}
If at least one of the off-diagonal blocks of the matrices $U^{(1)}$ and
$U^{(2)}$ (i.e.\ $U_{12}^{(1)}$, $U_{21}^{(1)}$, $U_{12}^{(2)}$, or
$U_{21}^{(2)}$) is of maximal rank, then the matrix $U^{(1)}$ is $V$-compatible
with $U^{(2)}$ for all unitary $p\times p$ matrices $V$.
\end{theorem}

\begin{proof}
We recall that for $p\leq n_1/2$ the $(n_1-p)\times p$ matrix $U_{12}^{(1)}$ is
not of maximal rank ($=\min\{n_1-p,p \}$) iff there is a vector $b\in\C^p$ such
that $U_{12}^{(1)}b=0$. For $p\geq n_1/2$ the matrix $U_{12}^{(1)}$ is not of
maximal rank iff there is a vector $c\in\C^p$ such that ${U_{12}^{(1)}}^\ast
c=0$.

Let us suppose that the matrix $U^{(1)}$ is not $V$-compatible with $U^{(2)}$.
Then by Lemma \ref{graph:neu:i:v} it follows that all off-diagonal blocks of
$U^{(1)}$ and $U^{(2)}$ are not of maximal rank.
\end{proof}

Actually we have also the following result. Let a unitary $n\times n$ matrix
$U$ be written in the block form
\begin{equation*}
U=\begin{pmatrix} U_{11} & U_{12} \\ U_{21} & U_{22} \end{pmatrix},
\end{equation*}
where $U_{11}$ is an $(n-p)\times(n-p)$ matrix, $U_{22}$ is a $p\times p$
matrix etc.\ with $p$ being an arbitrary integer such that $1\leq p < n$.

\begin{lemma}\label{lemma:3.8}
The matrices $U_{12}$ and $U_{21}$ are simultaneously either of maximal rank or
not of maximal rank.
\end{lemma}

\begin{proof}
Let us suppose that $p\leq n/2$. Then the $(n-p)\times p$ matrix $U_{12}$ is
not of maximal rank iff there is a non-zero vector $b\in\C^p$ such that
$U_{12}b=0$. From the unitarity of $U$ it follows that
\begin{equation*}
U_{11}^\ast U_{12}+U_{21}^\ast U_{22}=0,\qquad U_{22}^\ast U_{22}+ U_{12}^\ast
U_{12} =\1,
\end{equation*}
and therefore
\begin{equation*}
U_{21}^\ast U_{22}b=0,\qquad U_{22}^\ast U_{22}b=b.
\end{equation*}
Thus $U_{22}b\neq 0$ and $U_{22}b\in\Ker U_{21}^\ast$. From this it follows
that the $(n-p)\times p$ matrix $U_{21}^\ast$ is not of maximal rank, and thus
the $p\times(n-p)$ matrix $U_{21}$ is also not of maximal rank.

Now let us suppose that $n>p\geq n/2$. Then the matrix $U_{12}$ is not of
maximal rank iff there is a nontrivial vector $b\in\C^{n-p}$ such that
$U_{12}^\ast b=0$. Again from the unitarity we have
\begin{equation*}
U_{21}U_{11}^\ast + U_{22} U_{12}^\ast =0,\qquad U_{11}U_{11}^\ast+
U_{12}U_{12}^\ast=\1,
\end{equation*}
and therefore
\begin{equation*}
U_{21}U_{11}^\ast b=0,\qquad U_{11} U_{11}^\ast b=b.
\end{equation*}
Thus, the $p\times (n-p)$ matrix $U_{21}$ is not of maximal rank.
\end{proof}

We will show now that the $\ast$-product can be extended to arbitrary, not
necessarily $V$-compatible unitary matrices. We will prove that the operators
\begin{equation*}
U_{21}^{(2)} K_1 = U_{21}^{(2)}(\1-V U_{22}^{(1)}V^\ast U_{11}^{(2)})^{-1}V
\end{equation*}
and
\begin{equation*}
U_{12}^{(1)} K_2 = U_{12}^{(1)}(\1-V^\ast U_{11}^{(2)}V
U_{22}^{(1)})^{-1}V^\ast
\end{equation*}
are actually well-defined. From Lemma \ref{X.X.neu} it follows (see
\cite[Section I.5.3]{Kato}) that
\begin{equation*}
P_{\cC^\perp}(\1-V U_{22}^{(1)} V^\ast U_{11}^{(2)})^{-1}\quad\textrm{and}
\quad P_{\cB^\perp}(\1-V^\ast U_{11}^{(2)} V U_{22}^{(1)})^{-1}
\end{equation*}
are well-defined. From (ii) and (iii) of Lemma \ref{graph:neu:i:v} it follows
that $\cC\subseteq\Ker U_{21}^{(2)}$ and $\cB\subseteq\Ker U_{12}^{(1)}$ and
thus
\begin{equation*}
U_{21}^{(2)} (\1- V U_{22}^{(1)} V^\ast U_{11}^{(2)})^{-1} = U_{21}^{(2)}
P_{\cC^\perp}(\1-V U_{22}^{(1)} V^\ast U_{11}^{(2)})^{-1}
\end{equation*}
and
\begin{equation*}
U_{12}^{(1)}(\1-V^\ast U_{11}^{(2)} V U_{22}^{(1)})^{-1} =
U_{12}^{(1)}P_{\cB^\perp}(\1-V^\ast U_{11}^{(2)} V U_{22}^{(1)})^{-1}
\end{equation*}
are well-defined. Similarly one can show that the operators
\begin{equation*}
U_{21}^{(2)} V (\1- U_{22}^{(1)} V^\ast U_{11}^{(2)}
V)^{-1}\quad\textrm{and}\quad U_{12}^{(1)} V^\ast (\1- U_{11}^{(2)} V
U_{22}^{(1)} V^\ast)^{-1}
\end{equation*}
are also well-defined. With this we obtain that the relations \eqref{56} indeed
also define the generalized star product of two non-compatible unitary
matrices. Moreover, we have
\begin{equation}\label{Extra.Ref.Null}
\begin{split}
U_{21}^{(2)} (\1- V U_{22}^{(1)} V^\ast U_{11}^{(2)})^{-1} c = 0,& \qquad
U_{12}^{(1)}(\1-V^\ast U_{11}^{(2)} V U_{22}^{(1)})^{-1} b= 0,\\ U_{21}^{(2)} V
(\1- U_{22}^{(1)} V^\ast U_{11}^{(2)} V)^{-1} \widetilde{c} =0,& \qquad
U_{12}^{(1)} V^\ast (\1- U_{11}^{(2)} V U_{22}^{(1)} V^\ast)^{-1}
\widetilde{b}=0
\end{split}
\end{equation}
for all $c\in\cC$, $b\in\cB$, $\widetilde{c}\in\widetilde{\cC}$, and
$\widetilde{b}\in\widetilde{\cB}$.

\begin{theorem}\label{graph:th:4.1}
For arbitrary unitary matrices $U^{(1)}$, $U^{(2)}$, and $V$ the matrix
$U=U^{(1)}*_{V}U^{(2)}$ is unitary.
\end{theorem}

This theorem was proved in Appendix C of \cite{Kostrykin:Schrader:99b} in the
case when $U^{(1)}$ is $V$-compatible with $U^{(2)}$. For the general case the
proof is given in Appendix A below.

Analogously one can prove associativity of the generalized star product. More
precisely let $U^{(3)}$ be a unitary $n_3\times n_3$ and $V^\prime$ a unitary
$p^\prime\times p^\prime$ matrix with $p^\prime\leq n_2$, $p^\prime\leq n_3$.
If $p+p^\prime\leq n_1$, then
\begin{displaymath}
U^{(1)}*_{V}(U^{(2)}*_{V^{\prime}}U^{(3)})=
(U^{(1)}*_{V}U^{(2)})*_{V^{\prime}}U^{(3)}
\end{displaymath}
holds.

\begin{theorem}\label{Stetigkeit}
The generalized star product is a continuous operation in each of its two
arguments, i.e.\ for any unitary matrices $U^{(1)}$, $U^{(2)}$, $U^{(3)}$, and
$V$ such that $U^{(2)}$ and $U^{(3)}$ have equal size there is a constant $C>0$
depending on $U^{(1)}$ and $V$ only such that
\begin{eqnarray*}
\|U^{(1)}*_V U^{(2)}-U^{(1)}*_V U^{(3)}\|\leq C\|U^{(2)}-U^{(3)}\|,
\end{eqnarray*}
where the norm $\|\cdot\|$ is an arbitrary matrix norm. A similar estimate
holds with respect to the first argument.
\end{theorem}

In \cite{Kostrykin:Schrader:99b} we proved that the scattering matrix of a
self-adjoint Laplacian on an arbitrary graph is a continuous function of
$\lambda>0$. Theorem \ref{Stetigkeit} together with the composition rule given
in Section \ref{sec4} below allows to give an alternative proof of this fact.
We will not dwell on the details here.

In the sequel we will use the notion of the Moore-Penrose pseudoinverse (see
e.g.\ \cite{Strang}). Recall that for any (not necessarily square) matrix $M$
its pseudoinverse $M^\star$ is uniquely defined by the Penrose equations
\begin{eqnarray*}
M M^\star M = M, && M^\star M M^\star = M^\star,\\ (M^\star M)^\ast= M^\star M,
&& (M M^\star)^\ast = M M^\star.
\end{eqnarray*}
One also has
\begin{eqnarray*}
{M^\star}^\ast &=& {M^\ast}^\star,\\
\Ran M^\star &=& \Ran M^\ast,\\
\Ker M^\star &=& \Ker M^\ast,
\end{eqnarray*}
and $MM^\star=P_{\Ran M}$, $M^\star M=P_{\Ran M^\ast}$, where $P_{\cH}$ denotes
the orthogonal projection onto the linear subspace $\cH$. Moreover $0^\star=0$.
If $M$ is a square matrix of maximal rank then $M^\star=M^{-1}$.

Let $U$ again be an arbitrary unitary $n\times n$ matrix written in the block
form with some $1\leq p < n$.

\begin{lemma}\label{graph:hilfe}
If $\Ker U_{12}=\{0\}$ then $\Ker\left(U_{21}-U_{22} U_{12}^\star
U_{11}\right)^\ast=\{0\}$.
\end{lemma}

\begin{proof}
Assume the converse, i.e.\ let there be $c\in\C^p$, $c\neq 0$ such that
\begin{equation*}
\left(U_{21}-U_{22} U_{12}^\star
U_{11}\right)^\ast c=0,
\end{equation*}
or, equivalently,
\begin{equation*}
U_{21}^\ast c- U_{11}^\ast {U_{12}^\star}^\ast U_{22}^\ast c=0.
\end{equation*}
We multiply this equation by $U_{21}$ from the left and use the unitarity of
$U$ which in particular implies
\begin{equation*}
U_{21} U_{21}^\ast+U_{22} U_{22}^\ast=\1.
\end{equation*}
This yields
\begin{equation}\label{c.must.0}
c-U_{22}U_{22}^\ast c- U_{21}U_{11}^\ast {U_{12}^\star}^\ast U_{22}^\ast c=0.
\end{equation}
Again by unitarity we have $U_{21}U_{11}^\ast=-U_{22}U_{12}^\ast$. Recall that
\begin{equation*}
U_{12}^\ast {U_{12}^\star}^\ast = \left(U_{12}^\star U_{12}\right)^\ast
=(\1-P_{\Ker U_{12}})^\ast=\1
\end{equation*}
by the hypothesis of the lemma. Thus, from \eqref{c.must.0} it follows that
$c=0$.
\end{proof}

Now we turn to a discussion of the inverse of a unitary $2p\times 2p$ matrix
$U$ with respect to the generalized star product $*_p$, i.e.\ the existence of
the unitary matrices $U^L$ and $U^R$ such that
\begin{equation*}
U^L *_p\ U= U *_p\ U^R=\E,
\end{equation*}
where $\E$ is the $2p\times 2p$ matrix $\begin{pmatrix} 0 & \1\\
\1 & 0 \end{pmatrix}$ (in the $p\times p$ block notation). We will not discuss
general necessary and sufficient conditions for the existence of $U^L$ and
$U^R$, but simply restrict ourselves to a special case. We will prove

\begin{theorem}
Let $U$ be an arbitrary $2p\times 2p$ ($p\geq 1$) unitary matrix. Let at least
one of the $p\times p$ matrices $U_{12}$ and $U_{21}$ be of maximal rank
$(=p)$. Then there exists a unique unitary $2p\times 2p$ matrix $U^\prime$ such
that
\begin{equation}\label{graph:inverse}
U^\prime\ \ast_p\ U = U\ *_p\ U^\prime=\E.
\end{equation}
\end{theorem}

\begin{proof}
By Lemma \ref{lemma:3.8} both matrices $U_{12}$ and $U_{21}$ have maximal rank.
We will discuss only the second of the relations
\eqref{graph:inverse}. In block notation this relation has the form
\begin{equation}\label{graph:inverse:expand}
\begin{split}
U_{11} + U_{12} (\1-U_{11}^\prime U_{22})^{-1} U_{11}^\prime U_{21} =0,\\
U_{22}^\prime+ U_{21}^\prime(\1-U_{22}U_{11}^\prime)^{-1} U_{22}
U_{12}^\prime=0,\\ U_{12}(\1-U_{11}^\prime U_{22})^{-1}U_{12}^\prime=\1,\\
U_{21}^\prime(\1-U_{22}U_{11}^\prime)^{-1}U_{21}=\1.
\end{split}
\end{equation}
By Theorem \ref{graph:dosta} the matrix $U$ must be compatible with $U^\prime$
such that $(\1-U_{11}^\prime U_{22})^{-1}$ and $(\1-U_{22}U_{11}^\prime)^{-1}$
are both well defined. We multiply the first of the relations
\eqref{graph:inverse:expand} by $U_{12}^{-1}$ from the left. Next we multiply
the resulting expression by $\1-U_{11}^\prime U_{22}$ thus obtaining
\begin{equation*}
U_{11}^\prime (U_{21}-U_{22} U_{12}^{-1} U_{11})=-U_{12}^{-1}U_{11}.
\end{equation*}
By Lemma \ref{graph:hilfe} we have that $U_{21}-U_{22} U_{12}^{-1} U_{11}$ is
invertible and thus
\begin{equation}\label{u11.new}
U_{11}^\prime = -U_{12}^{-1} U_{11}(U_{21}-U_{22} U_{12}^{-1} U_{11})^{-1}.
\end{equation}
From the third relation in \eqref{graph:inverse:expand} we obtain
\begin{equation*}
U_{12}^\prime=(\1-U_{11}^\prime U_{22}) U_{12}^{-1} = U_{12}^{-1} + U_{12}^{-1}
U_{11}(U_{21}-U_{22} U_{12}^{-1} U_{11})^{-1} U_{22} U_{12}^{-1}.
\end{equation*}
The fourth relation in \eqref{graph:inverse:expand} gives
\begin{equation}\label{u21.new}
U_{21}^\prime=U_{21}^{-1}(\1-U_{22}U_{11}^\prime) = (U_{21}-U_{22} U_{12}^{-1}
U_{11})^{-1}.
\end{equation}
The second relation in \eqref{graph:inverse:expand} determines $U_{22}^\prime$.

It remains to prove that $U^\prime$ is unitary. By the unitarity of the matrix
$U$ we have
\begin{equation*}
(U_{21}^\ast - U_{11}^\ast {U_{12}^\ast}^{-1} U_{22}^\ast)(U_{21} - U_{22}
U_{12}^{-1} U_{11}) = \1 + U_{11}^\ast {U_{12}^\ast}^{-1} U_{12}^{-1} U_{11}.
\end{equation*}
Therefore
\begin{equation*}
(U_{21}^\ast - U_{11}^\ast {U_{12}^\ast}^{-1} U_{22}^\ast)^{-1} [\1 +
U_{11}^\ast {U_{12}^\ast}^{-1} U_{12}^{-1} U_{11}](U_{21} - U_{22} U_{12}^{-1}
U_{11})^{-1} =\1,
\end{equation*}
and thus by \eqref{u11.new} and \eqref{u21.new} we obtain
\begin{equation*}
{U_{11}^\prime}^\ast U_{11}^\prime + {U_{21}^\prime}^\ast U_{21}^\prime =\1.
\end{equation*}
The relations ${U_{12}^\prime}^\ast U_{12}^\prime + {U_{22}^\prime}^\ast
U_{22}^\prime =\1$, ${U_{11}^\prime}^\ast U_{12}^\prime + {U_{21}^\prime}^\ast
U_{22}^\prime =0$, and ${U_{12}^\prime}^\ast U_{11}^\prime +
{U_{22}^\prime}^\ast U_{21}^\prime =0$ can be proved similarly.

The left inverse is constructed similarly and by means of the obvious
relation
\begin{equation*}
(U_{21}-U_{22}U_{12}^{-1} U_{11})^{-1}U_{22}U_{12}^{-1} = U_{21}^{-1}U_{22}
(U_{21}-U_{22}U_{12}^{-1} U_{11})^{-1}U_{22}U_{12}^{-1}
\end{equation*}
is easily shown to be equal to the right inverse.
\end{proof}

\begin{corollary}
Let $\mathsf{G}$ be the set of all $2p\times 2p$ unitary matrices with $p\times
p$ blocks $U_{12}$ and $U_{21}$ both being of maximal rank, endowed with the
generalized star product $*_p$ as a multiplication and $\E$ as a unit. Then
$\mathsf{G}$ is a group isomorphic to $\U(p,p)$.
\end{corollary}

The proof will follow from the arguments given in Section \ref{sec5}. In
particular, the group isomorphism between $\mathsf{G}$ and $\U(p,p)$ is given
by the formulas \eqref{Lambda.form} and \eqref{S.repres} below. We note that
this isomorphism generalizes the well-known \emph{set} isomorphism between the
group $\SU(1,1)$ and a subgroup of $\SU(2)$.

The set of \emph{all} $2p\times 2p$ unitary matrices endowed with the
generalized star product $\ast_p$ as a multiplication and $\E$ as a unit is no
longer a group but only a semigroup.

\section{Composition Rule for the Scattering Matrices}\label{sec4}
\setcounter{equation}{0}

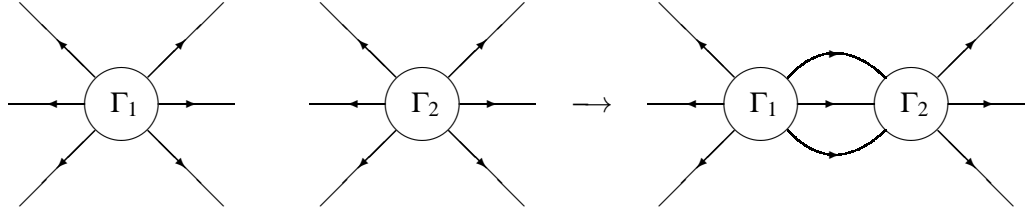
\begin{figure}[ht]
\centerline{
\unitlength1mm
\begin{picture}(140,60)
\put(20,30){\circle{10}}
\put(18.5,29){$\Gamma_1$}
\put(15,30){\line(-1,0){10}}
\put(25,30){\line(1,0){10}}
\put(16.465,33.535){\line(-1,1){10}}
\put(16.465,26.465){\line(-1,-1){10}}
\put(23.535,33.535){\line(1,1){10}}
\put(23.535,26.465){\line(1,-1){10}}
\put(15,30){\vector(-1,0){5}}
\put(16.465,33.535){\vector(-1,1){5}}
\put(16.465,26.465){\vector(-1,-1){5}}
\put(25,30){\vector(1,0){5}}
\put(23.535,33.535){\vector(1,1){5}}
\put(23.535,26.465){\vector(1,-1){5}}
\put(60,30){\circle{10}}
\put(58.5,29){$\Gamma_2$}
\put(55,30){\line(-1,0){10}}
\put(65,30){\line(1,0){10}}
\put(56.465,33.535){\line(-1,1){10}}
\put(56.465,26.465){\line(-1,-1){10}}
\put(63.535,33.535){\line(1,1){10}}
\put(63.535,26.465){\line(1,-1){10}}
\put(55,30){\vector(-1,0){5}}
\put(56.465,33.535){\vector(-1,1){5}}
\put(56.465,26.465){\vector(-1,-1){5}}
\put(65,30){\vector(1,0){5}}
\put(63.535,33.535){\vector(1,1){5}}
\put(63.535,26.465){\vector(1,-1){5}}
\put(80,29){$\longrightarrow$}
\put(105,30){\circle{10}}
\put(103.5,29){$\Gamma_1$}
\put(100,30){\line(-1,0){10}}
\put(110,30){\line(1,0){10}}
\put(101.465,33.535){\line(-1,1){10}}
\put(101.465,26.465){\line(-1,-1){10}}
\put(100,30){\vector(-1,0){5}}
\put(101.465,33.535){\vector(-1,1){5}}
\put(101.465,26.465){\vector(-1,-1){5}}
\put(110,30){\vector(1,0){5}}
\put(125,30){\circle{10}}
\put(123.5,29){$\Gamma_2$}
\put(130,30){\line(1,0){10}}
\put(128.535,33.535){\line(1,1){10}}
\put(128.535,26.465){\line(1,-1){10}}
\put(130,30){\vector(1,0){6}}
\put(128.535,33.535){\vector(1,1){5}}
\put(128.535,26.465){\vector(1,-1){5}}
\qbezier(108.535,33.535)(114.5,40)(121.465,33.535)
\qbezier(108.535,26.465)(114.5,20)(121.465,26.465)
\put(114.5,36.7){\vector(1,0){1}}
\put(114.5,23.2){\vector(1,0){1}}
\end{picture}}
\caption{\label{graph:fig:2} Gluing of two graphs.}
\end{figure}

Now we apply the generalized star product to prove the composition rule for the
scattering matrices on graphs. For this we only need the special case $V=\1_p$,
the $p\times p$ unit matrix, and so we introduce the notation
$*_{p}:=*_{V=\1_p}$. Let $\Gamma_1$ and $\Gamma_2$ be two graphs with $n_1\geq
1$ and $n_2\geq 1$ external lines, respectively, labeled by $\cE_1$ and
$\cE_2$, i.e.\ $\#(\cE_1)=n_1$, $\#(\cE_2)=n_2$ and an arbitrary number of
internal lines with given fixed lengths (see Fig.\ \ref{graph:fig:2}).
Furthermore at all vertices we have local boundary conditions giving Laplace
operators $\Delta(\Gamma_1)$ on $\Gamma_1$ and $\Delta(\Gamma_2)$ on $\Gamma_2$
and the scattering matrices $S_1(\lambda)$ and $S_2(\lambda)$. Let now
$\cE_1^{0}$ and $\cE_2^{0}$ be subsets of $\cE_1$ and $\cE_2$ respectively
having an equal number $1\leq p\leq
\min\{n_1,n_2\}$ of elements. Also let $\varphi_{0}:\;
\cE_1^{0}\rightarrow
\;\cE_2^{0}$ be a one-to-one map. Finally to each $k\in
\cE_1^{0}$ we associate a number $a_{k}>0$. With these data
we can now form a graph $\Gamma$ by connecting the external line
$k\in\cE_1^{0}$ with the line $\varphi_{0}(k)\in\cE_2^{0}$ to form a line of
length $a_{k}$. In other words any interval $[0_{k},\infty_{k})$, $k\in\cE_1^0$
belonging to $\Gamma_1$ and the interval
$[0_{\varphi_{0}(k)},\infty_{\varphi_{0}(k)})$ belonging to $\Gamma_2$ is
replaced by the finite interval $[0_{k},a_{k}]$ with $0_{k}$ being associated
to the same vertex in $\Gamma_1$ as previously and $a_{k}$ being associated to
the same vertex in $\Gamma_2$ as $0_{\varphi_{0}(k)}$ before in the sense of
the discussion at the end of the previous section. Recall that the graphs need
not be planar. Thus $\Gamma$ has $n=n_1+n_2-2p$ external lines indexed by
elements in $(\cE_1\setminus\cE_1^{0})\cup(\cE_2\setminus
\cE_2^{0})$ and $p$ internal lines indexed by elements in $\cE_1^{0}$
in addition to those of $\Gamma_1$ and of $\Gamma_2$. We denote this set by
$\cI_{12}$ such that the set of all internal lines of the graph $\Gamma$ is
given by $\cI=\cI_1\cup\cI_2\cup\cI_{12}$. There are no new vertices in
addition to those of $\Gamma_1$ and $\Gamma_2$ so the boundary conditions on
$\Gamma_1$ and $\Gamma_2$ define boundary conditions on $\Gamma$ resulting in a
Laplace operator $\Delta(\Gamma)$. Suppose that the indices of $\cE_1^{0}$ in
$\cE_1$ come after the indices in $\cE_1\setminus\cE_1^{0}$ (in an arbitrary
but fixed order) (see
\eqref{54}). Via the map $\varphi_{0}$ we may identify $\cE_2^{0}$
with $\cE_1^{0}$ so let these indices now come first in $\cE_2$, but again in
the same order. Finally, let the diagonal $n_2\times n_2$ matrix
$V(\underline{a})$ be given as
\begin{displaymath}
V(\underline{a})=\begin{pmatrix}
        \exp i\sqrt{\lambda}\underline{a}&0\\
        0&\1 \end{pmatrix},
\end{displaymath}
where $\exp(i\sqrt{\lambda}\underline{a})$ again is the diagonal $p\times p$
matrix given by the $p$ new lengths $a_{k}$, $k\in \cE_2^{0}$.

Recall that by Theorem \ref{graph:theorem:2} all eigenfunctions of the operator
$-\Delta(\Gamma)$ have the form
\begin{equation*}
\psi = \begin{cases} 0, & j\in \cE,\\ \alpha_j e^{i\sqrt{\lambda}x} +
\beta_j e^{-i\sqrt{\lambda}x}, & j\in\cI, \end{cases}
\end{equation*}
where the coefficients $\alpha_j$ and $\beta_j$ satisfy the homogeneous
equation
\begin{equation*}
Z_{A,B,\underline{a}}(\lambda)\begin{pmatrix} 0 \\ \alpha \\ \beta
\end{pmatrix} = 0,\qquad \alpha=\{\alpha_j\}_{j=1}^m,\qquad \beta=\{\beta_j\}_{j=1}^m
\end{equation*}
with the matrix $Z_{A,B,\underline{a}}(\lambda)$ defined by \eqref{Z.def}. We
define the linear subspace $\cL_{12}(\lambda)$ of $\C^{n+2m}$ as a set of all
vectors $(0, \alpha, \beta)^T$ for which $\alpha_j=\beta_j=0$ for all
$j\in\cI_{12}$,
\begin{equation}\label{L12.def}
\cL_{12}(\lambda) =\left\{\ell = (0,\alpha,\beta)^T\in\Ker
Z_{A,B,\underline{a}}(\lambda)\subset \C^{n+2m}\ \Big|\
\alpha_j=\beta_j\ \forall j\in\cI_{12}\right\}.
\end{equation}
Obviously for $\lambda\notin\sigma_{A,B,\underline{a}}$ this subspace is
trivial, i.e.\ $\cL_{12}(\lambda)=\{0\}$. Let
$\Upsilon(\Gamma,\cI_{12})\subset\R$ be the set of those eigenvalues of
$-\Delta(\Gamma)$ for which $\Ker Z_{A,B,\underline{a}}(\lambda)\ominus
\cL_{12}(\lambda)$ is nontrivial. Obviously the eigenfunctions corresponding
to the eigenvalues from $\Upsilon(\Gamma,\cI_{12})$ have nontrivial overlap
with $\cI_{12}$, i.e.\ $\supp\phi\cap \cI_{12}$ has non-zero measure.

Let $\Xi(\Gamma_1,\Gamma_2)\subset\R_+$ be the set of those $\lambda>0$ for
which
$\Ker(V(\underline{a})S_{22}^{(1)}(\lambda)V(\underline{a})S_{11}^{(2)}(\lambda)-1)$
is nontrivial.

\begin{theorem}\label{graph:thm:factorization}
With the above notations
$\Xi(\Gamma_1,\Gamma_2)=\Upsilon(\Gamma,\cI_{12})\cap\R_+$. The composition
rule
\begin{equation}\label{compos}
S(\lambda)=S_1(\lambda)*_p V(\underline{a})S_2(\lambda)V(\underline{a})
\end{equation}
holds for all $\lambda\in\R_+$. If $\lambda\in\Upsilon(\Gamma,\cI_{12})$ and
$\lambda>0$ then its multiplicity equals
\begin{eqnarray*}
\lefteqn{\dim\Ker(-\Delta(\Gamma_1)-\lambda)+\dim\Ker(-\Delta(\Gamma_2)-\lambda)}\\&&+
\dim\Ker(V(\underline{a})S_{22}^{(1)}(\lambda)V(\underline{a})S_{11}^{(2)}(\lambda)-1).
\end{eqnarray*}
In particular if the eigenvalue $\lambda>0$ is such that
$\lambda\notin\Upsilon(\Gamma,\cI_{12})$ then its multiplicity equals
\begin{equation*}
\dim\Ker(-\Delta(\Gamma_1)-\lambda)+\dim\Ker(-\Delta(\Gamma_2)-\lambda).
\end{equation*}
\end{theorem}

Here $\dim\Ker(-\Delta(\Gamma_l)-\lambda)$ denotes the multiplicity of the
eigenvalue $\lambda$ regardless whether it is embedded into the absolutely
continuous spectrum or not.

Note that by Lemma \ref{Frobenius}
$\dim\Ker(V(\underline{a})S_{22}^{(1)}(\lambda)V(\underline{a})
S_{11}^{(2)}(\lambda)-1)$ equals the algebraic multiplicity of the eigenvalue
$\mu=1$ of $V(\underline{a})S_{22}^{(1)}(\lambda)V(\underline{a})
S_{11}^{(2)}(\lambda)$.

If by cutting $p$ internal lines of an arbitrary graph $\Gamma$ with local
boundary conditions, the graph will be decomposed into two disjoint subgraphs
$\Gamma_1$ and $\Gamma_2$, by Theorem \ref{graph:thm:factorization} the
scattering matrix $S_\Gamma$ can be obtained from the scattering matrices
$S_{\Gamma_1}$ and $S_{\Gamma_2}$ at the same energy. Thus, using
\eqref{compos} iteratively the scattering matrix associated to any graph can be obtained
from the scattering matrices associated to its subgraphs each having one vertex
only. In fact, pick one vertex and choose all the internal lines connecting to
all other vertices. This leads to two graphs and the rule
\eqref{compos} may be applied. Iterating this procedure $L$ times, where $L$ is
the number of vertices, gives the desired result.

\begin{proof}[Proof of Theorem \ref{graph:thm:factorization}]
We split the proof into several steps.

1. First we suppose that $S_1(\lambda)$ is compatible with
$V(\underline{a})S_2(\lambda)V(\underline{a})$ and prove that the composition
rule \eqref{compos} holds. Let $\psi_j^k(x,\lambda;\Gamma_l)$, $j\in
\cE_l\cup\cI_l$ for any $k\in\cE_l$ be the solution of the Schr\"{o}dinger equation
with the operator $-\Delta(\Gamma_l)$, $l=1,2$ (see
\eqref{10}) at energy $\lambda$. Let $\Psi_l(x,\lambda)$ be $n_l\times n_l$ matrices
\begin{equation}\label{psi.l}
\left[\Psi_l(x,\lambda) \right]_{jk}=\psi_j^k(x,\lambda;\Gamma_l), \quad j,k\in\cE_l,
\quad l=1,2
\end{equation}
such that
\begin{equation}\label{graph:standard}
\Psi_l(x,\lambda)=e^{-i\sqrt{\lambda}x}\1+e^{i\sqrt{\lambda}x}S_l(\lambda),\quad l=1,2.
\end{equation}
Observe that $e^{-i\sqrt{\lambda}x}$ and $e^{i\sqrt{\lambda}x}$ are linearly
independent functions and therefore the scattering matrices may uniquely be
recovered from $\Psi_l(x,\lambda)$. The columns of $\Psi_l(x,\lambda)$ define
the external part of solutions of the Schr\"{o}dinger equation for
$-\Delta(\Gamma_l)$ at energy $\lambda$. We are looking for a square matrix
\begin{displaymath}
\Psi(x,\lambda)=e^{-i\sqrt{\lambda}x}\1 + e^{i\sqrt{\lambda}x} S(\lambda)
\end{displaymath}
such that its $(n_1-p)+(n_2-p)=n_1+n_2-2p$ columns defines the external parts
of a solution to the Schr\"{o}dinger equation for $-\Delta(\Gamma)$. Here the
indices are assumed to be arranged such that the first indices are those of
$\cE_1\setminus\cE_1^0$ followed by the indices of $\cE_2\setminus\cE_2^0$. The
aim is to obtain $\Psi(x,\lambda)$ from $\Psi_1(x,\lambda)$,
$\Psi_2(x,\lambda)$, and the lengths $\underline{a}=\{a_s\}_{s\in\cI_{12}}$ of
the new internal lines $\cI_{12}$. By the above observation this will determine
the scattering matrix $S(\lambda)$. The strategy will be to find new solutions
of the Schr\"{o}dinger equations for $-\Delta(\Gamma_l)$ with incoming plane waves
($e^{-i\sqrt{\lambda}x}$) in the channels $k\in\cE_l\setminus\cE_l^0$ which
agree suitably in the channels $k\in\cE_1^0$ and $\varphi_0(k)\in\cE_2^0$.

With the conventions made above we write
\begin{equation}\label{S1.S2}
\begin{split}
S_1(\lambda) &= \begin{pmatrix}
    S_1^{(n_1-p)\times(n_1-p)}(\lambda) & S_1^{(n_1-p)\times p}(\lambda) \\
    S_1^{p\times(n_1-p)}(\lambda) & S_1^{p\times p}(\lambda)
    \end{pmatrix},\\ S_2(\lambda) &= \begin{pmatrix}
    S_2^{p\times p}(\lambda) & S_2^{p\times (n_2-p)}(\lambda) \\
    S_2^{(n_2-p)\times p}(\lambda) & S_2^{(n_2-p)\times (n_2-p)}(\lambda)
    \end{pmatrix},\\
\end{split}
\end{equation}
where the superscripts denote the sizes of the blocks. For arbitrary
$p\times(n_1-p)$ matrices $C_1$ and $C_2$, respectively, consider the
$n_1\times (n_1-p)$ and $n_2\times(n_1-p)$ matrices
\begin{equation}\label{Phi1.Phi2}
\begin{split}
\Phi_1(x,\lambda;C_1) = e^{-i\sqrt{\lambda}x}\begin{pmatrix}
           \1 \\ C_1 \end{pmatrix}
           +e^{i\sqrt{\lambda}x}S_1(\lambda)\begin{pmatrix}
           \1 \\ C_1 \end{pmatrix},\\
\Phi_2(x,\lambda;C_2) = e^{-i\sqrt{\lambda}x}\begin{pmatrix}
           C_2 \\ 0 \end{pmatrix}
           +e^{i\sqrt{\lambda}x}S_2(\lambda)\begin{pmatrix}
           C_2 \\ 0 \end{pmatrix}.\\
\end{split}
\end{equation}
Here $\1$ stands for the $(n_1-p)\times(n_1-p)$ unit matrix and $0$ stands for
the $(n_2-p)\times(n_1-p)$ zero matrix. Obviously, the columns of
$\Phi_l(x,\lambda;C_l)$ are the external parts of linear combinations of the
columns of $\Psi_l(x,\lambda;\Gamma_l)$, and thus define the external parts of
solutions of the Schr\"{o}dinger equations for the operators $-\Delta(\Gamma_l)$,
$l=1,2$. Note that $\Phi_1(x,\lambda;C_1)$ has an incoming plane wave in any of
the channels $k\in\cE_1\setminus\cE_1^0$ and $\Phi_2(x,\lambda;C_2)$ has no
incoming plane wave in all channels $k\in\cE_2\setminus\cE_2^0$.

Now we make the coordinate transformations $x\rightarrow a_k-x$ on the lines
$\varphi_0(k)\in\cE_2^0$ ($k\in\cE_1^0$). The reason for this is as follows.
Under the gluing process $\Gamma_1, \Gamma_2\rightarrow\Gamma$ (see Fig.\
\ref{graph:fig:2}) the two half-lines corresponding to $k\in\cE_1^0$ and
$\varphi_0(k)\in\cE_2^0$ are replaced by the interval $[0,a_k]$, giving the new
lines in $\cI_{12}$. This may be realized by identifying a point $P$ on the
half-line corresponding to $k\in\cE_1^0$ and with coordinate $x$ ($0\leq x\leq
a_k$) with the point $Q$ on the half-line corresponding to
$\varphi_0(k)\in\cE_2^0$ with coordinate $a_k-x$. In particular $x=a_k$ on the
$k$-line corresponds to $x=0$ on the $\varphi_0(k)$-line and vice versa.
Applying this transformation to $\Phi_2(x,\lambda;C_2)$ we obtain in this new
coordinate system
\begin{eqnarray*}
\Phi_2^{(\underline{a})}(x,\lambda;C_2)=\begin{pmatrix}
        e^{-i\sqrt{\lambda}(\underline{a}-x)} & 0 \\
        0 & e^{-i\sqrt{\lambda} x}\end{pmatrix} \begin{pmatrix}
                C_2 \\ 0 \end{pmatrix} +
        \begin{pmatrix}
        e^{i\sqrt{\lambda}(\underline{a}-x)} & 0 \\
        0 & e^{i\sqrt{\lambda} x} \end{pmatrix} S_2(\lambda) \begin{pmatrix}
                C_2 \\ 0 \end{pmatrix}.
\end{eqnarray*}
We now require that $\Phi_1(x,\lambda;C_1)$ and
$\Phi_2^{(\underline{a})}(x,\lambda;C_2)$ agree on the lines labeled by
$\cI_{12}$. This will fix $C_1$ and $C_2$. Indeed, we then obtain
\begin{eqnarray*}
e^{-i\sqrt{\lambda}x} C_1 + e^{i\sqrt{\lambda}x} S_1^{p\times (n_1-p)}(\lambda)
+e^{i\sqrt{\lambda}x} S_1^{p\times p}(\lambda) C_1\\
=e^{-i\sqrt{\lambda}(\underline{a}-x)} C_2 + e^{i\sqrt{\lambda}(\underline{a}-x)}
S_2^{p\times p}(\lambda) C_2.
\end{eqnarray*}
By the linear independence of the functions $e^{i\sqrt{\lambda}x}$ and
$e^{-i\sqrt{\lambda}x}$ it follows that
\begin{equation}\label{inhom.1}
\begin{split}
& C_1 = e^{i\sqrt{\lambda}\underline{a}} S_2^{p\times p}(\lambda) C_2,\\
 & S_1^{p\times p}(\lambda) C_1 + S_1^{p\times (n_1-p)}(\lambda) =
 e^{-i\sqrt{\lambda}\underline{a}} C_2,\\
\end{split}
\end{equation}
and thus
\begin{equation}\label{C1.C2}
\begin{split}
C_2 &= \left[\1-e^{i\sqrt{\lambda}\underline{a}}S_1^{p\times
p}(\lambda)e^{i\sqrt{\lambda}\underline{a}} S_2^{p\times p}(\lambda)
\right]^{-1}e^{i\sqrt{\lambda}\underline{a}} S_1^{p\times (n_1-p)}(\lambda),\\
C_1 &= e^{i\sqrt{\lambda}\underline{a}} S_2^{p\times p}(\lambda)
\left[\1-e^{i\sqrt{\lambda}\underline{a}}S_1^{p\times
p}(\lambda)e^{i\sqrt{\lambda}\underline{a}} S_2^{p\times p}(\lambda)
\right]^{-1}e^{i\sqrt{\lambda}\underline{a}} S_1^{p\times (n_1-p)}(\lambda).\\
\end{split}
\end{equation}
Since for any invertible $A$ and $U$ the identities $UA^{-1}=(AU^{-1})^{-1}$
and $A^{-1}U=(U^{-1}A)^{-1}$ hold, we have
\begin{eqnarray*}
\lefteqn{\left[\1-e^{i\sqrt{\lambda}\underline{a}}S_1^{p\times
p}(\lambda)e^{i\sqrt{\lambda}\underline{a}} S_2^{p\times p}(\lambda)
\right]^{-1}e^{i\sqrt{\lambda}\underline{a}}}\\
&=&
\left[e^{-i\sqrt{\lambda}\underline{a}}- S_1^{p\times
p}(\lambda)e^{i\sqrt{\lambda}\underline{a}} S_2^{p\times p}(\lambda)
\right]^{-1}\\
&=& e^{i\sqrt{\lambda}\underline{a}}\left[\1-S_1^{p\times
p}(\lambda)e^{i\sqrt{\lambda}\underline{a}} S_2^{p\times p}(\lambda)
e^{i\sqrt{\lambda}\underline{a}} \right]^{-1}.
\end{eqnarray*}
Since $S_1(\lambda)$ is assumed to be compatible with
$V(\underline{a})S_2(\lambda)V(\underline{a})$ the inverses in \eqref{C1.C2}
are well defined.

Similar to \eqref{S1.S2} and according to the ordering convention made above we
write the scattering matrix $S(\lambda)$ for the graph $\Gamma$ in the block
form
\begin{displaymath}
S(\lambda) = \begin{pmatrix}
     S^{(n_1-p)\times(n_1-p)}(\lambda) & S^{(n_1-p)\times(n_2-p)}(\lambda)\\
     S^{(n_2-p)\times(n_1-p)}(\lambda) & S^{(n_2-p)\times(n_2-p)}(\lambda)
     \end{pmatrix},
\end{displaymath}
where the superscripts denote again the sizes of the blocks. Since
$\Phi_1(x,\lambda;C_1)$ has an incoming plane wave in any of the first $n_1-p$
channels $k\in\cE_1\setminus\cE_1^0$, equations \eqref{C1.C2} allow one to
determine $S^{(n_1-p)\times(n_1-p)}(\lambda)$ and
$S^{(n_1-p)\times(n_2-p)}(\lambda)$:
\begin{eqnarray*}
S^{(n_1-p)\times(n_1-p)}(\lambda) &=& S_1^{(n_1-p)\times(n_1-p)}(\lambda)
  +S_1^{(n_1-p)\times p}(\lambda)C_1\\
  &=& S_1^{(n_1-p)\times(n_1-p)}(\lambda) +
  S_1^{(n_1-p)\times p}(\lambda)e^{i\sqrt{\lambda}\underline{a}}
  S_2^{p\times p}(\lambda) e^{i\sqrt{\lambda}\underline{a}}\\
  &&\quad\cdot\left[\1-S_1^{p\times p}(\lambda)e^{i\sqrt{\lambda}\underline{a}}
  S_2^{p\times p}(\lambda)e^{i\sqrt{\lambda}\underline{a}} \right]^{-1}
  S_1^{p\times(n_1-p)}(\lambda),\\
S^{(n_2-p)\times(n_1-p)}(\lambda) &=& S_2^{(n_2-p)\times p}(\lambda)C_2\\
&=&S_2^{(n_2-p)\times p}(\lambda)e^{i\sqrt{\lambda}\underline{a}}\\
  &&\quad\cdot
  \left[\1-S_1^{p\times p}(\lambda)e^{i\sqrt{\lambda}\underline{a}}
  S_2^{p\times p}(\lambda)e^{i\sqrt{\lambda}\underline{a}} \right]^{-1}
  S_1^{p\times(n_1-p)}(\lambda).
\end{eqnarray*}

To determine the remaining blocks of the scattering matrix $S(\lambda)$ instead
of \eqref{Phi1.Phi2} we consider the $n_1\times(n_2-p)$ and $n_2\times(n_2-p)$
matrices
\begin{equation}\label{Phi.tilde.1.Phi.tilde.2}
\begin{split}
\widetilde{\Phi}_1(x,\lambda;\widetilde{C}_1) = e^{-i\sqrt{\lambda}x}\begin{pmatrix}
           0 \\ \widetilde{C}_1 \end{pmatrix}
           +e^{i\sqrt{\lambda}x}S_1(\lambda)\begin{pmatrix}
           0 \\ \widetilde{C}_1 \end{pmatrix},\\
\widetilde{\Phi}_2(x,\lambda;\widetilde{C}_2) = e^{-i\sqrt{\lambda}x}\begin{pmatrix}
           \widetilde{C}_2 \\ \1 \end{pmatrix}
           +e^{i\sqrt{\lambda}x}S_2(\lambda)\begin{pmatrix}
           \widetilde{C}_2 \\ \1 \end{pmatrix}.
\end{split}
\end{equation}
with arbitrary $p\times(n_2-p)$ matrices $\widetilde{C}_1$ and
$\widetilde{C}_2$. Again $\widetilde{\Phi}_l(x,\lambda;\widetilde{C}_l)$ are
the external parts of some solutions of the Schr\"{o}dinger equations with the
operators $-\Delta(\Gamma_l)$, $l=1,2$. Now
$\widetilde{\Phi}_1(x,\lambda;\widetilde{C}_1)$ has no incoming plane waves in
the channels $k\in\cE_1\setminus\cE_1^0$, but
$\widetilde{\Phi}_2(x,\lambda;\widetilde{C}_2)$ has an incoming plane wave in
any of the channels $k\in\cE_2\setminus\cE_2^0$.

Repeating the arguments used above we obtain the following matching conditions
for $\widetilde{C}_1$ and $\widetilde{C}_2$:
\begin{equation}\label{inhom.2}
\begin{split}
& \widetilde{C}_2 = e^{i\sqrt{\lambda}\underline{a}}S_1^{p\times
p}\widetilde{C}_1,\\ & S_2^{p\times p}(\lambda)\widetilde{C}_2 +
S_2^{p\times(n_2-p)}(\lambda)=e^{-i\sqrt{\lambda}\underline{a}}\widetilde{C}_1,\\
\end{split}
\end{equation}
and thus
\begin{eqnarray*}
\widetilde{C}_1 &=& \left[\1-e^{i\sqrt{\lambda}\underline{a}}S_2^{p\times p}(\lambda)
e^{i\sqrt{\lambda}\underline{a}}S_1^{p\times p}(\lambda)
\right]^{-1}e^{i\sqrt{\lambda}\underline{a}} S_2^{p\times(n_2-p)}(\lambda),\\
\widetilde{C}_2 &=& e^{i\sqrt{\lambda}\underline{a}}S_1^{p\times p}(\lambda)
\left[\1-e^{i\sqrt{\lambda}\underline{a}}S_2^{p\times p}(\lambda)
e^{i\sqrt{\lambda}\underline{a}}S_1^{p\times p}(\lambda)
\right]^{-1}e^{i\sqrt{\lambda}\underline{a}} S_2^{p\times(n_2-p)}(\lambda).
\end{eqnarray*}
Since $S_1(\lambda)$ is compatible with
$V(\underline{a})S_2(\lambda)V(\underline{a})$ the inverses are again well
defined. From this and from
\eqref{Phi.tilde.1.Phi.tilde.2} it follows that
\begin{eqnarray*}
S^{(n_2-p)\times (n_2-p)}(\lambda) &=& S_2^{(n_2-p)\times (n_2-p)}(\lambda) +
     S_2^{(n_2-p)\times p}(\lambda)\widetilde{C}_2\\
     &=& S_2^{(n_2-p)\times (n_2-p)}(\lambda) + S_2^{(n_2-p)\times p}(\lambda)
     e^{i\sqrt{\lambda}\underline{a}}S_1^{p\times p}(\lambda)\\
     &&\quad\cdot
     \left[\1-e^{i\sqrt{\lambda}\underline{a}}S_2^{p\times p}(\lambda)
e^{i\sqrt{\lambda}\underline{a}}S_1^{p\times p}(\lambda)
\right]^{-1}e^{i\sqrt{\lambda}\underline{a}} S_2^{p\times (n_2-p)}(\lambda),\\
S^{(n_1-p)\times (n_2-p)}(\lambda) &=& S_1^{(n_1-p)\times
  p}(\lambda)\widetilde{C}_1\\ &=& S_1^{(n_1-p)\times p}(\lambda)\\
  &&\quad\cdot\left[\1-e^{i\sqrt{\lambda}\underline{a}}S_2^{p\times p}(\lambda)
e^{i\sqrt{\lambda}\underline{a}}S_1^{p\times p}(\lambda)
\right]^{-1}e^{i\sqrt{\lambda}\underline{a}} S_2^{p\times (n_2-p)}(\lambda).
\end{eqnarray*}
By the definition of the generalized star product \eqref{56} we obtain
\eqref{compos}.

2. Now suppose that $\lambda\in\Xi(\Gamma_1,\Gamma_2)$. We prove that the
composition rule \eqref{compos} remains valid. Also $\lambda\in\Upsilon(\Gamma,
\cI_{12})$ and the multipicity of $\lambda$ equals
\begin{eqnarray}\label{mult.ziel}
\lefteqn{\dim\Ker(-\Delta(\Gamma_1)-\lambda)+\dim\Ker(-\Delta(\Gamma_2)-\lambda)}\nonumber\\&&+
\dim\Ker(V(\underline{a})S_{22}^{(1)}(\lambda)V(\underline{a})S_{11}^{(2)}(\lambda)-1).
\end{eqnarray}

The assumption $\lambda\in\Xi(\Gamma_1,\Gamma_2)$ implies that
\begin{displaymath}
\1-S_1^{p\times p}(\lambda)e^{i\sqrt{\lambda}\underline{a}}S_2^{p\times
p}(\lambda)e^{i\sqrt{\lambda}\underline{a}}
\end{displaymath}
and
\begin{displaymath}
\1-e^{i\sqrt{\lambda}\underline{a}}S_2^{p\times
p}(\lambda)e^{i\sqrt{\lambda}\underline{a}}S_1^{p\times p}(\lambda)
\end{displaymath}
have nontrivial kernels. This implies that the homogeneous form of the
equations \eqref{inhom.1} and \eqref{inhom.2},
\begin{equation}\label{sol.1}
\begin{split}
& C_1=e^{i\sqrt{\lambda}\underline{a}} S_2^{p\times p}(\lambda)C_2,\\ &
S_1^{p\times p}(\lambda)C_1= e^{-i\sqrt{\lambda}\underline{a}}C_2,\\
\end{split}
\end{equation}
and
\begin{equation}\label{sol.2}
\begin{split}
& \widetilde{C}_2= e^{i\sqrt{\lambda}\underline{a}} S_1^{p\times
p}(\lambda)\widetilde{C}_1,\\ & S_2^{p\times
p}(\lambda)\widetilde{C}_2=e^{-i\sqrt{\lambda}\underline{a}}\widetilde{C}_1,\\
\end{split}
\end{equation}
respectively, have nontrivial solutions. It is easy to prove that the
inhomogeneous equations \eqref{inhom.1} and \eqref{inhom.2} still have
solutions in this case. Consider for instance the equation \eqref{inhom.1},
which is equivalent to
\begin{displaymath}
C_2 = e^{i\sqrt{\lambda}\underline{a}}
S_1^{p\times(n_1-p)}(\lambda)+e^{i\sqrt{\lambda}\underline{a}}S_1^{p\times
p}(\lambda) e^{i\sqrt{\lambda}\underline{a}}S_2^{p\times p}(\lambda) C_2.
\end{displaymath}
By the Fredholm alternative this equation has a non-trivial solution iff
\begin{equation}\label{viele.punkte}
S_1^{p\times(n_1-p)}(\lambda)^\ast e^{-i\sqrt{\lambda}\underline{a}}b=0
\end{equation}
for any $0\neq b\in\C^{p}$ satisfying
\begin{displaymath}
S_2^{p\times p}(\lambda)^\ast e^{-i\sqrt{\lambda}\underline{a}} S_1^{p\times
p}(\lambda)^\ast e^{-i\sqrt{\lambda}\underline{a}} b = b.
\end{displaymath}
By Lemma \ref{graph:lem:factor} with $A=S_2^{p\times p}(\lambda)^\ast$ and
$B=e^{-i\sqrt{\lambda}\underline{a}}S_1^{p\times p}(\lambda)^\ast
e^{-i\sqrt{\lambda}\underline{a}}$ we have
\begin{equation}\label{bbbbb}
e^{i\sqrt{\lambda}\underline{a}} S_1^{p\times p}(\lambda)S_1^{p\times
p}(\lambda)^\ast e^{-i\sqrt{\lambda}\underline{a}}b=b.
\end{equation}
From the unitarity of $V(\underline{a})S_1(\lambda)V(\underline{a})$, which
states in particular that
\begin{eqnarray*}
\lefteqn{e^{i\sqrt{\lambda}\underline{a}} S_1^{p\times p}(\lambda)S_1^{p\times
p}(\lambda)^\ast e^{-i\sqrt{\lambda}\underline{a}}}\\ &&+
e^{i\sqrt{\lambda}\underline{a}}S_1^{p\times(n_1-p)}(\lambda)
S_1^{p\times(n_1-p)}(\lambda)^\ast e^{-i\sqrt{\lambda}\underline{a}}= \1,
\end{eqnarray*}
and from \eqref{bbbbb} it follows that
\begin{displaymath}
e^{i\sqrt{\lambda}\underline{a}}S_1^{p\times(n_1-p)}(\lambda)
S_1^{p\times(n_1-p)}(\lambda)^\ast e^{-i\sqrt{\lambda}\underline{a}}b=0.
\end{displaymath}
Since $\Ker C^\ast C=\Ker C$ for any operator $C$ we obtain
\eqref{viele.punkte}. Equation \eqref{inhom.2} is discussed similarly.

From \eqref{Phi1.Phi2} and \eqref{Phi.tilde.1.Phi.tilde.2} it follows that the
Schr\"{o}dinger equation with the operator $-\Delta(\Gamma)$ for given value of the
spectral parameter $\lambda>0$ has (nonunique) solutions which have the form
\begin{equation}\label{zzz.1}
\begin{split}
\Phi_1(x,\lambda;C_1) = e^{-i\sqrt{\lambda}x}\begin{pmatrix}
           0 \\ C_1 \end{pmatrix}
           +e^{i\sqrt{\lambda}x}S_1(\lambda)\begin{pmatrix}
           0 \\ C_1 \end{pmatrix},\\
\Phi_2(x,\lambda;C_2) = e^{-i\sqrt{\lambda}x}\begin{pmatrix}
           C_2 \\ 0 \end{pmatrix}
           +e^{i\sqrt{\lambda}x}S_2(\lambda)\begin{pmatrix}
           C_2 \\ 0 \end{pmatrix}
\end{split}
\end{equation}
and
\begin{equation}\label{zzz.2}
\begin{split}
\widetilde{\Phi}_1(x,\lambda;\widetilde{C}_1) = e^{-i\sqrt{\lambda}x}\begin{pmatrix}
           0 \\ \widetilde{C}_1 \end{pmatrix}
           +e^{i\sqrt{\lambda}x}S_1(\lambda)\begin{pmatrix}
           0 \\ \widetilde{C}_1 \end{pmatrix},\\
\widetilde{\Phi}_2(x,\lambda;\widetilde{C}_2) = e^{-i\sqrt{\lambda}x}\begin{pmatrix}
           \widetilde{C}_2 \\ 0 \end{pmatrix}
           +e^{i\sqrt{\lambda}x}S_2(\lambda)\begin{pmatrix}
           \widetilde{C}_2 \\ 0 \end{pmatrix},
\end{split}
\end{equation}
where $C_1$ and $C_2$ ($\widetilde{C}_1$ and $\widetilde{C}_2$) solve
\eqref{sol.1} (\eqref{sol.2}, respectively). Note that $C_1=\widetilde{C}_1$
and $C_2=\widetilde{C}_2$. On the lines in the set $\cI_{12}$ the quantity
$\Phi_1$ coincides with $\Phi_2$ and $\widetilde{\Phi}_1$ and
$\widetilde{\Phi}_2$. We will now prove that
\begin{equation}\label{zzz.3}
S_1^{(n_1-p)\times p}(\lambda) C_1=0,\qquad S_2^{(n_2-p)\times p}(\lambda)
C_2=0.
\end{equation}
Thus, the functions \eqref{zzz.1} and \eqref{zzz.2} are zero on all external
lines of the graph $\Gamma$ and their support has nontrivial overlap with the
interval lines $\cI_{12}$.

From \eqref{sol.1} it follows that $C_1=e^{i\sqrt{\lambda}\underline{a}}
S_2^{p\times p}(\lambda)e^{i\sqrt{\lambda}\underline{a}}S_1^{p\times
p}(\lambda)C_1$. By Lemma \ref{graph:lem:factor} we have
\begin{equation*}
S_1^{p\times p}(\lambda)^\ast S_1^{p\times p}(\lambda)C_1 =C_1.
\end{equation*}
By unitarity it follows that
\begin{equation*}
S_1^{(n_1-p)\times p}(\lambda)^\ast S_1^{(n_1-p)\times p}(\lambda) C_1=0
\end{equation*}
and thus $S_1^{(n_1-p)\times p}(\lambda) C_1=0$. The second relation in
\eqref{zzz.3} is proved similarly.

Now we note that from \eqref{sol.1} and \eqref{sol.2} it follows that
\begin{equation*}
\Rank C_1 = \Rank C_2 = \dim\Ker
\left(e^{i\sqrt{\lambda}\underline{a}} S_2^{p\times p}(\lambda)
e^{i\sqrt{\lambda}\underline{a}} S_1^{p\times p}(\lambda)-1\right).
\end{equation*}
The columns of \eqref{zzz.1} correspond to linear independent eigenfunctions of
$-\Delta(\Gamma)$ for the eigenvalue $\lambda$. There are precisely $\dim\Ker
\left(e^{i\sqrt{\lambda}\underline{a}} S_2^{p\times p}(\lambda)
e^{i\sqrt{\lambda}\underline{a}} S_1^{p\times p}(\lambda)-1\right)$ such
eigenfunctions and the supports of all them have nontrivial overlap with the
internal lines $\cI_{12}$.

3. Let $\lambda\in\Upsilon(\Gamma,\cI_{12})\cap \R_+$ and let
\begin{equation}\label{Doppelstern}
\dim\left(\Ker Z_{A,B,\underline{a}}(\lambda)\ominus \cL_{12}(\lambda)\right) = k,
\end{equation}
where the linear subspace $\cL_{12}(\lambda)$ is defined by \eqref{L12.def}.
This means that there are precisely $k$ eigenfunctions of $-\Delta(\Gamma)$
which disappear if we cut the internal lines $\cI_{12}$. We will prove that
$\lambda\in\Xi(\Gamma_1,\Gamma_2)$ and that
\begin{equation}\label{Tripplestern}
\dim\Ker(V(\underline{a})S_{22}^{(1)}(\lambda)V(\underline{a})S_{11}^{(2)}(\lambda)-1)=k
\end{equation}
which in turn implies that
\begin{eqnarray*}
\dim\Ker(-\Delta(\Gamma)-\lambda)=\dim\Ker(-\Delta(\Gamma_1)-\lambda)
+\dim\Ker(-\Delta(\Gamma_2)-\lambda)\\+\dim\Ker
(V(\underline{a})S_{22}^{(1)}(\lambda)V(\underline{a})S_{11}^{(2)}(\lambda)-1).
\end{eqnarray*}
From the existence of the above mentioned eigenfunctions it follows that these
eigenfunctions can be constructed by means of superposition and matching of the
solutions \eqref{graph:standard} of the Schr\"{o}dinger equation for the operators
$-\Delta(\Gamma_1)$ and $-\Delta(\Gamma_2)$ at energy $\lambda>0$. For any
vectors $C_1, C_2\in\C^p$ the functions
\begin{eqnarray*}
\phi_1(x,\lambda,C_1) = e^{-i\sqrt{\lambda}x}\begin{pmatrix} 0 \\ C_1 \end{pmatrix} +
e^{i\sqrt{\lambda}x} S_1(\lambda)\begin{pmatrix} 0 \\ C_1 \end{pmatrix},\\
\phi_2(x,\lambda,C_2) = e^{-i\sqrt{\lambda}x}\begin{pmatrix} C_2 \\ 0 \end{pmatrix} +
e^{i\sqrt{\lambda}x} S_2(\lambda)\begin{pmatrix} C_2 \\ 0 \end{pmatrix}
\end{eqnarray*}
define the external parts of solutions of the Schr\"{o}dinger equations for the
operators $-\Delta(\Gamma_l)$, $l=1,2$. Since the eigenfunctions are supported
on internal lines of the graph $\Gamma$ (Theorem \ref{graph:theorem:2}) the
vectors  $C_1$ and $C_2$ must satisfy
\begin{equation*}
S_1^{(n_1-p)\times p}(\lambda) C_1 =0,\qquad S_2^{p\times (n_2-p)}(\lambda) C_2
=0
\end{equation*}
such that $\phi_1(x,\lambda,C_1)$ vanishes in any of the channels
$k\in\cE_1\setminus\cE_1^0$ and $\phi_2(x,\lambda,C_2)$ vanishes in all
channels $k\in\cE_2\setminus\cE_2^0$. Making the coordinate transformation
$x\rightarrow\underline{a}-x$ on the lines $\varphi_0(k)\in\cE_2^0\
(k\in\cE_1^0)$ and requiring that $\phi_1(x,\lambda,C_1)$ and
$\phi_2(\underline{a}-x,\lambda,C_2)$ agree on the lines labeled by $\cI_{12}$,
we obtain
\begin{eqnarray*}
C_1 = e^{i\sqrt{\lambda}\underline{a}} S_2^{p\times p}(\lambda) C_2,\\
S_1^{p\times p}(\lambda) C_1 = e^{-i\sqrt{\lambda}\underline{a}} C_2,
\end{eqnarray*}
or equivalently
\begin{equation}\label{Singlestern}
\begin{split}
& e^{i\sqrt{\lambda}\underline{a}} S_2^{p\times p}(\lambda)
e^{i\sqrt{\lambda}\underline{a}} S_1^{p\times p}(\lambda) C_1 = C_1,\\ & C_2 =
e^{i\sqrt{\lambda}\underline{a}} S_1^{p\times p}(\lambda) C_1.
\end{split}
\end{equation}
Linear independent solutions of \eqref{Singlestern} correspond to linear
independent eigenfunctions of $-\Delta(\Gamma)$ and vise versa. Thus the
condition \eqref{Doppelstern} implies \eqref{Tripplestern}. This completes the
proof of the theorem.
\end{proof}

Note that if $\Gamma$ is simply the disjoint union of $\Gamma_1$ and
$\Gamma_2$, i.e.\ if no connections are made (corresponding to $p=0$ and
$n=n_1+n_2$), then $S(\lambda)$ is just the direct sum of $S_1(\lambda)$ and
$S_2(\lambda)$. Also $V^{\ast}S(\lambda)V=S_{2n}^\mathrm{free}(\lambda)*_V
S(\lambda)$ for any scattering matrix with $n$ open ends and any unitary
$n\times n$ matrix $V$, where
\begin{displaymath}
S_{2n}^\mathrm{free}(\lambda)=\begin{pmatrix}
                                    0 & \1 \\
                                    \1 & 0
                                    \end{pmatrix}.
\end{displaymath}
Similarly $S(\lambda)*_V S_{2n}^\mathrm{free}(\lambda)=VS(\lambda)V^{\ast}$.

\begin{figure}[ht]
\centerline{
\unitlength1mm
\begin{picture}(120,40)
\put(20,20){\line(1,0){80}}
\put(50,20){\circle*{2}}
\put(70,20){\circle*{2}}
\put(40,20){\vector(-1,0){10}}
\put(50,20){\vector(1,0){10}}
\put(80,20){\vector(1,0){10}}
\put(40,21){1}
\put(60,21){2}
\put(80,21){3}
\end{picture}}
\caption{\label{graph:fig:line} The graph from Example \ref{graph:ex:5}.}
\end{figure}
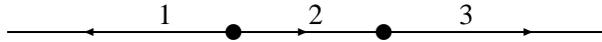

\begin{example}\label{graph:ex:5}
Consider an arbitrary self-adjoint Laplacian $\Delta(A,B)$ with local boundary
conditions on the graph depicted in Fig.\ \ref{graph:fig:line}, where the
distance between the two vertices is $a$. The composition rule (\ref{compos})
with
\begin{displaymath}
V(a)=\begin{pmatrix} e^{i\sqrt{\lambda}a} & 0 \\ 0 & 1
\end{pmatrix}
\end{displaymath}
easily gives
\begin{equation}\label{spec.case}
\begin{aligned}
S_{11} &= S^{(1)}_{11}+S^{(1)}_{12}S_{11}^{(2)}S^{(1)}_{21}
(1-S^{(1)}_{22}S^{(2)}_{11}e^{2ia\sqrt{\lambda}})^{-1},\\ S_{22}
&=S^{(2)}_{22}+ S_{22}^{(1)} S_{21}^{(2)}S_{12}^{(2)}
(1-S^{(1)}_{22}S^{(2)}_{11}e^{2ia\sqrt{\lambda}})^{-1},\\ S_{12} &=
S_{12}^{(1)}
S_{12}^{(2)}(1-S^{(1)}_{22}S^{(2)}_{11}e^{2ia\sqrt{\lambda}})^{-1},\\ S_{21} &=
S_{21}^{(2)}S_{21}^{(1)}(1-S^{(1)}_{22}S^{(2)}_{11}e^{2ia\sqrt{\lambda}})^{-1},
\end{aligned}
\end{equation}
where the S-matrices are written in the form analogous to \eqref{54}
\begin{displaymath}
S^{(1)} = \begin{pmatrix}S^{(1)}_{11} & S^{(1)}_{12} \\
                    S^{(1)}_{21} & S_{22}^{(1)}\end{pmatrix},\qquad
S^{(2)} = \begin{pmatrix}S^{(2)}_{11} & S^{(2)}_{12}\\
                    S^{(2)}_{21} & S_{22}^{(2)}\end{pmatrix},
\end{displaymath}
leaving out the $\lambda-$dependence. These relations are equivalent to the
well-known factorization formula
\cite{Redheffer:61,Redheffer:62,Kowal,Aktosun,Rozman,Bianchi:Ventra:95,Bianchi:Ventra:95a}
applied to the Laplacian on a line with boundary conditions posed at $x=0$ and
$x=a$.
\end{example}

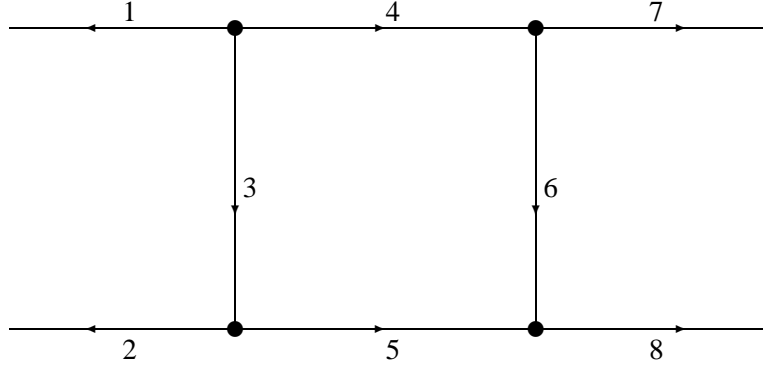
\begin{figure}[ht]
\centerline{
\unitlength1mm
\begin{picture}(120,60)
\put(10,50){\line(1,0){100}}
\put(10,10){\line(1,0){100}}
\put(40,50){\circle*{2}}
\put(40,10){\circle*{2}}
\put(80,50){\circle*{2}}
\put(80,10){\circle*{2}}
\put(40,50){\line(0,-1){40}}
\put(80,50){\line(0,-1){40}}
\put(30,50){\vector(-1,0){10}}
\put(30,10){\vector(-1,0){10}}
\put(90,50){\vector(1,0){10}}
\put(90,10){\vector(1,0){10}}
\put(25,51){1}
\put(25,6){2}
\put(95,51){7}
\put(95,6){8}
\put(40,35){\vector(0,-1){10}}
\put(80,35){\vector(0,-1){10}}
\put(41,27.5){3}
\put(81,27.5){6}
\put(50,50){\vector(1,0){10}}
\put(50,10){\vector(1,0){10}}
\put(60,51){4}
\put(60,6){5}
\end{picture}}
\caption{\label{graph:fig:neu} The graph from Example \ref{graph:ex:6}.
The arrows show the positive direction for every edge. The edges 3 and 6 have
the length $a$ and the edges 4 and 5 the length $b$.}
\end{figure}

\begin{example}\label{graph:ex:6}
Consider the graph depicted in Fig.\ \ref{graph:fig:neu} where the length of
the edges 3 and 6 equals $a$ and the length of the edges 4 and 5 equals $b$.
Let the boundary conditions be given as
\begin{equation}\label{randbed}
\begin{split}
\psi_1(0)=\psi_3(0)=\psi_4(0),\\
\psi_2(0)=\psi_3(a)=\psi_5(0),\\
\psi_1^\prime(0)+\psi_3^\prime(0)+\psi_4^\prime(0)=0,\\
\psi_2^\prime(0)+\psi_5^\prime(0)-\psi_3^\prime(a)=0,
\end{split}
\end{equation}
\begin{eqnarray*}
\psi_4(b)=\psi_6(0)=\psi_7(0),\\
\psi_5(b)=\psi_6(a)=\psi_8(0),\\
-\psi_4^\prime(b)+\psi_6^\prime(0)+\psi_7^\prime(0)=0,\\
-\psi_5^\prime(b)-\psi_6^\prime(a)+\psi_8^\prime(0)=0.
\end{eqnarray*}
Obviously they define a self-adjoint operator which we denote by $\Delta(a,b)$.
The scattering matrix corresponding to this operator (as defined by
\eqref{S.matrix:2spaces} and \eqref{10}) will be denoted by $S_{a,b}(\lambda)$.
To determine this $4\times 4$ matrix we first consider the graph depicted in
Fig.\ \ref{graph:fig:1} where the length of the edge 3 is supposed to be equal
$a$. The boundary conditions \eqref{randbed} determine the self-adjoint
operator. The corresponding scattering matrix we denote by $S_a(\lambda)$.
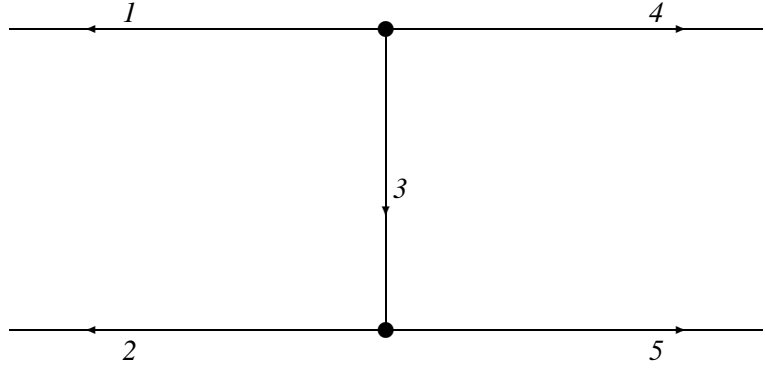
\begin{figure}[ht]
\centerline{
\unitlength1mm
\begin{picture}(120,60)
\put(10,50){\line(1,0){100}}
\put(10,10){\line(1,0){100}}
\put(60,50){\circle*{2}}
\put(60,10){\circle*{2}}
\put(60,50){\line(0,-1){40}}
\put(30,50){\vector(-1,0){10}}
\put(30,10){\vector(-1,0){10}}
\put(90,50){\vector(1,0){10}}
\put(90,10){\vector(1,0){10}}
\put(25,51){1}
\put(25,6){2}
\put(95,51){4}
\put(95,6){5}
\put(60,35){\vector(0,-1){10}}
\put(61,27.5){3}
\end{picture}}
\caption{\label{graph:fig:1} The graph from Example \ref{graph:ex:6}.
The arrows show the positive direction for every edge.}
\end{figure}
From \eqref{cont} it follows that
\begin{eqnarray}\label{scatt.mat.4.3}
S_a(\lambda)=\left(e^{i\sqrt{\lambda}a}-9e^{-i\sqrt{\lambda}a}\right)^{-1}
\qquad\qquad\qquad\qquad\qquad\qquad\qquad\qquad\qquad\\
{\cdot\scriptscriptstyle\begin{pmatrix}
e^{i\sqrt{\lambda}a}+3e^{-i\sqrt{\lambda}a} & -4
 & 2(e^{i\sqrt{\lambda}a}-3e^{-i\sqrt{\lambda}a}) & -4 \\
-4 &
e^{i\sqrt{\lambda}a}+3e^{-i\sqrt{\lambda}a} & -4 &
2(e^{i\sqrt{\lambda}a}-3e^{-i\sqrt{\lambda}a}) \\
 2(e^{i\sqrt{\lambda}a}-3e^{-i\sqrt{\lambda}a})& -4 &
 e^{i\sqrt{\lambda}a}+3e^{-i\sqrt{\lambda}a} & -4
\\
-4 &  2(e^{i\sqrt{\lambda}a}-3e^{-i\sqrt{\lambda}a}) & -4 &
e^{i\sqrt{\lambda}a}+3e^{-i\sqrt{\lambda}a} \end{pmatrix}}.\nonumber
\end{eqnarray}
By Theorem \ref{graph:thm:factorization} the scattering matrix
$S_{a,b}(\lambda)$ is given by
\begin{equation}\label{decomp:rule}
S_{a,b}(\lambda) = S_a(\lambda) *_2 V(\underline{b}) S_a(\lambda)
V(\underline{b}),
\end{equation}
where
\begin{displaymath}
V(\underline{b})=\diag(e^{i\sqrt{\lambda}b},e^{i\sqrt{\lambda}b},1,1),\qquad
\underline{b}=(b,b)\in\R^2.
\end{displaymath}

We now compute the $2\times 2$ matrices $K_1$ and $K_2$ entering the definition
\eqref{56} of the generalized star product, thus obtaining
\begin{eqnarray*}
&& K_1^{-1}=K_2^{-1}
=\left(e^{i\sqrt{\lambda}a}-9e^{-i\sqrt{\lambda}a}\right)^{-2} L,\\
&& (L)_{11}=(L)_{22}= e^{2i\sqrt{\lambda}a}(1-e^{2i\sqrt{\lambda}b}) +9
e^{-2i\sqrt{\lambda}a} (9-e^{2i\sqrt{\lambda}b}) -2(9+ 11
e^{2i\sqrt{\lambda}b}),\\ && (L)_{12}=(L)_{21}=
8e^{2i\sqrt{\lambda}b}(e^{i\sqrt{\lambda}a}+3e^{-i\sqrt{\lambda}a}).
\end{eqnarray*}
From this it follows that
\begin{eqnarray*}
\lefteqn{\det K_1^{-1}
=\det K_2^{-1} = \left(e^{i\sqrt{\lambda}a}-9e^{-i\sqrt{\lambda}a}\right)^{-4}}\\
&&\cdot e^{-4i\sqrt{\lambda}a}
\Big[\xi(\xi\eta^2-64)(\xi-8)^2+16\eta(-256-128\xi+44\xi^2-3\xi^3)\Big],
\end{eqnarray*}
where $\xi=\exp\{2i\sqrt{\lambda}a\}-1$ and $\eta=\exp\{2i\sqrt{\lambda}b\}-1$.
Obviously these determinants vanish if
$e^{2i\sqrt{\lambda}a}=e^{2i\sqrt{\lambda}b}=1$. One can show that there are no
other zeros. Note that the embedded eigenvalues of the operator $-\Delta(a,b)$
are determined by the equation $e^{2i\sqrt{\lambda}a}=e^{2i\sqrt{\lambda}b}=1$
such that for incommensurable $a$ and $b$ there are no embedded eigenvalues.

For $e^{2i\sqrt{\lambda}a}=e^{2i\sqrt{\lambda}b}=1$ the matrix $S_a(\lambda)$
is not compatible with $V(\underline{b})S_a(\lambda)V(\underline{b})$ and
\begin{displaymath}
K_1^{-1}=K_2^{-1}=\frac{1}{2}
\begin{pmatrix} 1 & \pm 1 \\ \pm 1 & 1 \end{pmatrix},
\end{displaymath}
where $\pm 1$ corresponds to $\exp\{i\sqrt{\lambda}a\}=\pm 1$. Obviously $\Ker
K_1^{-1}=\Ker K_2^{-1}$ is the subspace spanned by the vector $(1,\mp 1)^T$.
Further,
\begin{displaymath}
(S_a(\lambda))_{12}\begin{pmatrix} 1 \\ \mp 1 \end{pmatrix} =\frac{1}{2}
\begin{pmatrix} 1 & \pm 1 \\ \pm 1 & 1 \end{pmatrix}
\begin{pmatrix} 1 \\ \mp 1 \end{pmatrix}=0
\end{displaymath}
and
\begin{displaymath}
(V(\underline{b})S_a(\lambda)V(\underline{b}))_{12}\begin{pmatrix} 1 \\ \mp 1
\end{pmatrix} = \frac{1}{2}e^{i\sqrt{\lambda}b}
\begin{pmatrix} 1 & \pm 1 \\ \pm 1 & 1 \end{pmatrix}
\begin{pmatrix} 1 \\ \mp 1 \end{pmatrix}=0.
\end{displaymath}
Thus, as proved in Section \ref{sec3}, the generalized star product is well
defined also in the case when the matrix $S_a(\lambda)$ is not compatible with
$V(\underline{b})S_a(\lambda)V(\underline{b})$.
\end{example}

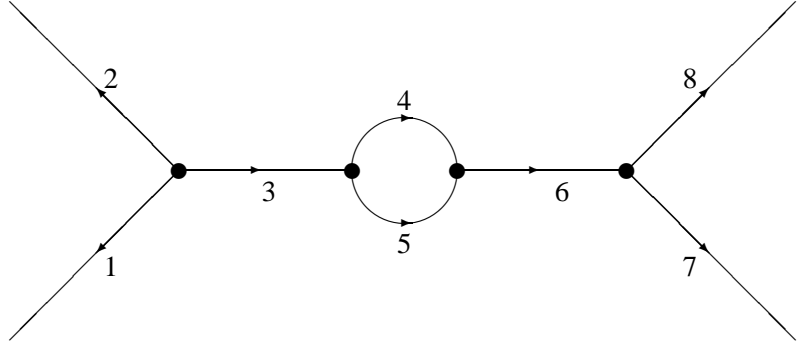
\begin{figure}[ht]
\centerline{
\unitlength1mm
\begin{picture}(120,60)
\put(60,30){\circle{15}}
\put(30,30){\line(1,0){22.5}}
\put(30,30){\vector(1,0){11}}
\put(30,30){\circle*{2}}
\put(30,30){\line(-1,1){22.5}}
\put(30,30){\vector(-1,1){11}}
\put(30,30){\line(-1,-1){22.5}}
\put(30,30){\vector(-1,-1){11}}
\put(53,30){\circle*{2}}
\put(67,30){\circle*{2}}
\put(67,30){\line(1,0){22.5}}
\put(67,30){\vector(1,0){11}}
\put(89.5,30){\circle*{2}}
\put(89.5,30){\line(1,1){22.5}}
\put(89.5,30){\vector(1,1){11}}
\put(89.5,30){\line(1,-1){22.5}}
\put(89.5,30){\vector(1,-1){11}}
\put(60,37){\vector(1,0){1}}
\put(60,23){\vector(1,0){1}}
\put(20,41){2}
\put(20,16){1}
\put(97,41){8}
\put(97,16){7}
\put(41,26){3}
\put(80,26){6}
\put(59,38){4}
\put(59,19){5}
\end{picture}}
\caption{\label{graph:fig:neu1} The graph from Example \ref{graph:ex:6}.
The arrows show the positive direction for every edge. The edges 3 and 6 have
the length $a$ and the edges 4 and 5 the length $b$.}
\end{figure}

\begin{example}\label{graph:ex:7}
Consider the graph depicted in Fig.\ \ref{graph:fig:neu1} where the length of
the edges 3 and 6 equals $a$ and the length of the edges 4 and 5 equals $b$.
Let the boundary conditions be given by
\begin{eqnarray*}
\psi_1(0)=\psi_2(0)=\psi_3(0),\\
\psi_4(0)=\psi_5(0)=\psi_3(a),\\
\psi_1^\prime(0)+\psi_2^\prime(0)+\psi_3^\prime(0)=0,\\
\psi_4^\prime(0)+\psi_5^\prime(0)-\psi_3^\prime(a)=0,\\
\psi_4(b)=\psi_5(b)=\psi_6(0),\\
\psi_6(a)=\psi_7(0)=\psi_8(0),\\
-\psi_4^\prime(b)-\psi_5^\prime(b)+\psi_6^\prime(0)=0,\\
-\psi_6^\prime(a)+\psi_7^\prime(0)+\psi_8^\prime(0)=0.
\end{eqnarray*}
Obviously they define a self-adjoint operator which we denote by $\Delta(a,b)$.
The scattering matrix corresponding to this operator (as defined by
\eqref{S.matrix:2spaces} and \eqref{10}) will be denoted by $S_{a,b}(\lambda)$.
To determine this $4\times 4$ matrix we first consider the graph depicted in
Fig.\ \ref{graph:fig:1} where the length of the edge 3 is supposed to be equal
$a$. The boundary conditions \eqref{randbed} determine the self-adjoint
operator.
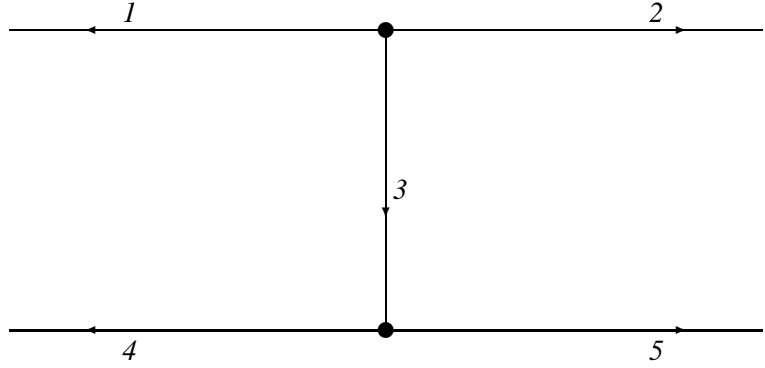
\begin{figure}[ht]
\centerline{
\unitlength1mm
\begin{picture}(120,60)
\put(10,50){\line(1,0){100}}
\put(10,10){\line(1,0){100}}
\put(60,50){\circle*{2}}
\put(60,10){\circle*{2}}
\put(60,50){\line(0,-1){40}}
\put(30,50){\vector(-1,0){10}}
\put(30,10){\vector(-1,0){10}}
\put(90,50){\vector(1,0){10}}
\put(90,10){\vector(1,0){10}}
\put(25,51){1}
\put(25,6){4}
\put(95,51){2}
\put(95,6){5}
\put(60,35){\vector(0,-1){10}}
\put(61,27.5){3}
\end{picture}}
\caption{\label{graph:fig:1a} The same graph as in Fig.\ \ref{graph:fig:1}
but with different ordering of the external lines.}
\end{figure}
The corresponding scattering matrix $S_a(\lambda)$ can be obtained from
\eqref{scatt.mat.4.3} by means of the permutation of its lines and columns thus
giving
\begin{eqnarray*}
S_a(\lambda)=\left(e^{i\sqrt{\lambda}a}-9e^{-i\sqrt{\lambda}a}\right)^{-1}
\qquad\qquad\qquad\qquad\qquad\qquad\qquad\qquad\qquad\\
\cdot{\scriptscriptstyle\begin{pmatrix} e^{i\sqrt{\lambda}a}+3e^{-i\sqrt{\lambda}a}
& 2(e^{i\sqrt{\lambda}a}-3e^{-i\sqrt{\lambda}a}) & -4 & -4 \\
2(e^{i\sqrt{\lambda}a}-3e^{-i\sqrt{\lambda}a}) &
e^{i\sqrt{\lambda}a}+3e^{-i\sqrt{\lambda}a} & -4 & -4 \\
-4 & -4 & e^{i\sqrt{\lambda}a}+3e^{-i\sqrt{\lambda}a} &
2(e^{i\sqrt{\lambda}a}-3e^{-i\sqrt{\lambda}a})\\
-4 & -4 & 2(e^{i\sqrt{\lambda}a}-3e^{-i\sqrt{\lambda}a}) &
e^{i\sqrt{\lambda}a}+3e^{-i\sqrt{\lambda}a} \end{pmatrix}}.
\end{eqnarray*}
By Theorem \ref{graph:thm:factorization} the scattering matrix
$S_{a,b}(\lambda)$ is given by
\begin{equation}\label{decomp:rule1}
S_{a,b}(\lambda) = S_a(\lambda) *_2 V(\underline{b}) S_a(\lambda)
V(\underline{b}),
\end{equation}
where
\begin{displaymath}
V(\underline{b})=\diag(e^{i\sqrt{\lambda}b},e^{i\sqrt{\lambda}b},1,1),\qquad
\underline{b}=(b,b).
\end{displaymath}

We now compute the $2\times 2$ matrices $K_1$ and $K_2$ entering the definition
\eqref{56} of the generalized star product, thus obtaining
\begin{eqnarray*}
&& K_1^{-1}=K_2^{-1}
=\left(e^{i\sqrt{\lambda}a}-9e^{-i\sqrt{\lambda}a}\right)^{-2} L,\\
&& (L)_{11}=(L)_{22}= e^{2i\sqrt{\lambda}a}(1-5 e^{2i\sqrt{\lambda}b}) +9
e^{-2i\sqrt{\lambda}a} (9-5 e^{2i\sqrt{\lambda}b}) -18(1-
e^{2i\sqrt{\lambda}b}),\\ && (L)_{12}=(L)_{21}=
-4e^{2i\sqrt{\lambda}b}(e^{2i\sqrt{\lambda}a}-9e^{-2i\sqrt{\lambda}a}).
\end{eqnarray*}
From this it follows that
\begin{eqnarray*}
\det K_1^{-1} =\det K_2^{-1} = \left(e^{i\sqrt{\lambda}a}-9e^{-i\sqrt{\lambda}a}\right)^{-4}
\Big[e^{4i\sqrt{\lambda}a}(1-10e^{2i\sqrt{\lambda}b} + 9e^{4i\sqrt{\lambda}b})\\
+ 9^3e^{-4i\sqrt{\lambda}a}
(9-10e^{2i\sqrt{\lambda}b}+e^{4i\sqrt{\lambda}b})-36 e^{2i\sqrt{\lambda}a}
(1-6e^{2i\sqrt{\lambda}b}+5e^{4i\sqrt{\lambda}b})\\ -18^2
e^{-2i\sqrt{\lambda}a} (9-14 e^{2i\sqrt{\lambda}b}+5e^{4i\sqrt{\lambda}b}) +18
(27
-86e^{2i\sqrt{\lambda}b} +59 e^{4i\sqrt{\lambda}b})
\Big].
\end{eqnarray*}
Obviously these determinants vanish if $e^{2i\sqrt{\lambda}b}=1$. One can show
that there are no other zeros. Note that the embedded eigenvalues of the
operator $-\Delta(a,b)$ are determined by the equation
$e^{2i\sqrt{\lambda}b}=1$.

For $e^{2i\sqrt{\lambda}b}=1$ the matrices $S_a(\lambda)$ and
$V(\underline{b})S_a(\lambda)V(\underline{b})$ are not compatible and
\begin{displaymath}
K_1^{-1}=K_2^{-1}=-4\frac{e^{2i\sqrt{\lambda}a}-9e^{-2i\sqrt{\lambda}a}}
{(e^{i\sqrt{\lambda}a}-9e^{-i\sqrt{\lambda}a})^2}
\begin{pmatrix} 1 & 1 \\ 1& 1 \end{pmatrix}.
\end{displaymath}
Obviously $\Ker K_1^{-1}=\Ker K_2^{-1}$ is the subspace spanned by the vector
$(1,-1)^T$. Further,
\begin{displaymath}
(S_a(\lambda))_{12}\begin{pmatrix} 1 \\ -1 \end{pmatrix} =-4
\begin{pmatrix} 1 & 1 \\ 1 & 1 \end{pmatrix}
\begin{pmatrix} 1 \\ -1 \end{pmatrix}=0
\end{displaymath}
and
\begin{displaymath}
(V(\underline{b})S_a(\lambda)V(\underline{b}))_{12}\begin{pmatrix} 1 \\ -1
\end{pmatrix} =-4 e^{i\sqrt{\lambda}b}
\begin{pmatrix} 1 & 1 \\ 1 & 1 \end{pmatrix}
\begin{pmatrix} 1 \\ -1 \end{pmatrix}=0.
\end{displaymath}
Thus, as proved in Section \ref{sec3}, the generalized star product is well
defined also in the case when the matrices $S_a(\lambda)$ and
$V(\underline{b})S_a(\lambda)V(\underline{b})$ are not compatible.

\end{example}

As already discussed in \cite{Kostrykin:Schrader:99b} multiple application of
\eqref{compos} to an arbitrary graph allows one by complete induction on the
number of vertices to calculate its scattering matrix from the scattering
matrices corresponding to single-vertex graphs. If these single vertex graphs
contain no tadpoles, i.e.\ internal lines starting and ending at the same
vertex, then \eqref{compos} give a complete explicit construction of the
scattering matrix in terms of the scattering matrices for single vertex graphs.
In case when a resulting single-vertex graph contains tadpoles we proceed as
follows. Let the graph $\Gamma$ have one vertex, $n$ external lines and $m$
tadpoles of lengths $a_i$. To calculate the scattering matrix of $\Gamma$ we
insert an extra vertex on each of the internal lines (for definiteness, say, at
$x=a_{i}/2$). At these new vertices we impose trivial boundary conditions
corresponding to continuous differentiability at this point. With these new
vertices we may now repeat our previous procedure. Thus in the end we arrive at
graphs with one vertex only and no tadpoles.

\section{Special Case $n_1=n_2=2p$: Transfer Matrices}\label{sec5}
\setcounter{equation}{0}

This section is devoted to the construction of the transfer matrix for
Schr\"{o}dinger operators on graphs with an even number of external lines. The
transfer matrix formalism for general Schr\"{o}dinger operators on the line is well
known (see e.g.\ \cite{Cycon:Froese:Kirsch:Simon}). Its relation to the
scattering matrix is discussed in e.g.\ \cite{Tong,Kostrykin:Schrader:99a}.

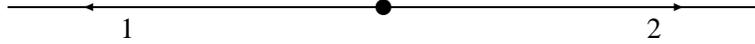
\begin{figure}[ht]
\centerline{
\unitlength1mm
\begin{picture}(120,20)
\put(10,10){\line(1,0){100}}
\put(60,10){\circle*{2}}
\put(30,10){\vector(-1,0){10}}
\put(90,10){\vector(1,0){10}}
\put(25,6){1}
\put(95,6){2}
\end{picture}}
\caption{\label{graph:point} The graph with $n=2$ and $m=0$.}
\end{figure}

We start with the simplest example of a Laplace operator on the graph with
$n=2$ and $m=0$ (see Fig.\ \ref{graph:point}) which is equivalent to a
Schr\"{o}dinger operator on the line with point interaction. The boundary
conditions given by the relation
\begin{equation}\label{graph:rb}
\begin{pmatrix}\psi_2(0) \\ \psi_2^\prime(0)\end{pmatrix} =
e^{i\mu}\begin{pmatrix} a & b \\ c & d \end{pmatrix}
\begin{pmatrix}\psi_1(0) \\ -\psi_1^\prime(0)\end{pmatrix},
\end{equation}
where the matrix
\begin{equation*}
\begin{pmatrix} a & b \\ c & d \end{pmatrix}\in \SL(2,\R),
\end{equation*}
and $\mu$ is real, lead to self-adjoint Laplacians (see
\cite{Kostrykin:Schrader:99b,Seba:86a,Carreau,Kurasov:96,Albeverio:Dabrowski:Kurasov}).
Conversely, from the viewpoint of the von Neumann extension theory (see e.g.\
\cite{RS2}) relation \eqref{graph:rb} describes almost all (with respect to the
Haar measure on $\U(2)$) self-adjoint Laplacians $\Delta(A,B)$. If
$\exp\{2i\mu\}=1$ the operator $\Delta(A,B)$ is real, i.e.\ commutes with
complex conjugation. In particular, the choice $a-1=d-1=b=0$, $\exp\{2i\mu\}=1$
corresponds to the $\delta$-potential of strength $c$ (see e.g.\
\cite{Albeverio:book}).

By definition the transfer matrix is a $2\times 2$ matrix
$M(\lambda)\in\U(1)\times\SL(2,\R)$ satisfying
\begin{equation}\label{transfer:1}
M(\lambda)\begin{pmatrix}\psi_1(0) \\ \psi_1^\prime(0)\end{pmatrix} =
\begin{pmatrix}\psi_2(0) \\ -\psi_2^\prime(0)\end{pmatrix}.
\end{equation}
In fact, in the case at hand it is given explicitly as follows
\begin{equation*}
M(\lambda)=e^{i\mu}\begin{pmatrix} a & b \\ c & d \end{pmatrix}
\end{equation*}
If $\exp\{2i\mu\}=1$ the matrix $M(\lambda)$ is unimodular, i.e.\
$M(\lambda)\in\SL(2;\R)$.

The transfer matrix possesses the following equivalent description. Any
solution of the Schr\"{o}dinger equation with the operator $-\Delta(A,B)$ for the
energy $\lambda>0$ has the form
\begin{eqnarray*}
u_1(x)=a_1 e^{i\sqrt{\lambda}x} + b_1 e^{-i\sqrt{\lambda}x},\\ u_2(x)=a_2
e^{i\sqrt{\lambda}x} + b_2 e^{-i\sqrt{\lambda}x}.
\end{eqnarray*}
From this and \eqref{transfer:1} it follows that there is a matrix
$\Lambda(\lambda)\in\U(1)\times\SU(1,1)\subset\U(1)\times\SL(2;\C)$ (with the
inclusion in the group-theoretical sense) such that
\begin{equation}\label{graph:property}
\Lambda(\lambda)\begin{pmatrix} a_1 \\ b_1 \end{pmatrix}=
\begin{pmatrix} b_2 \\ a_2 \end{pmatrix}
\end{equation}
and
\begin{equation*}
M(\lambda)=\begin{pmatrix} 1 & 1 \\ i\sqrt{\lambda} & -i\sqrt{\lambda}
\end{pmatrix} \Lambda(\lambda) \begin{pmatrix} 1 & 1 \\ i\sqrt{\lambda} & -i\sqrt{\lambda}
\end{pmatrix}^{-1}.
\end{equation*}
For $\lambda>0$ the matrix $\Lambda(\lambda)$ is related to the scattering
matrix
\begin{equation*}
S_{A,B}(\lambda)=\begin{pmatrix} R(\lambda) & T_1(\lambda) \\ T_2(\lambda) &
L(\lambda)
\end{pmatrix}
\end{equation*}
by the relation
\begin{equation*}
\Lambda(\lambda)=\begin{pmatrix}\frac{\displaystyle 1}{\displaystyle T_2(\lambda)} &
     -\frac{\displaystyle R(\lambda)}{\displaystyle T_2(\lambda)} \\[2ex]
\frac{\displaystyle L(\lambda)}{\displaystyle T_2(\lambda)} &
\frac{\displaystyle T_1(\lambda)}{\displaystyle |T_2(\lambda)|^2} \end{pmatrix},
\end{equation*}
where
\begin{eqnarray*}
T_1(\lambda) &=& 2e^{i\mu}(a-ib\sqrt{\lambda}+ic/\sqrt{\lambda}+d)^{-1},\\
T_2(\lambda) &=& 2e^{-i\mu}(a-ib\sqrt{\lambda}+ic/\sqrt{\lambda}+d)^{-1},\\
R(\lambda) &=&
(a-ib\sqrt{\lambda}+ic/\sqrt{\lambda}+d)^{-1}(a-ib\sqrt{\lambda}-ic/\sqrt{\lambda}-d),\\
L(\lambda) &=&
(a-ib\sqrt{\lambda}+ic/\sqrt{\lambda}+d)^{-1}(-a-ib\sqrt{\lambda}-ic/\sqrt{\lambda}+d).
\end{eqnarray*}
Note that $T_1(\lambda)=T_2(\lambda)$ for all $\lambda>0$ if the operator
$\Delta(A,B)$ is real, i.e.\ $\exp\{2i\mu\}=1$. This is in analogy with
Schr\"{o}dinger operators on the line with potentials which are necessarily real
(see e.g. \cite{Faddeev,Deift:Trubowitz}).

The factorization rule from Example \ref{graph:ex:5} can now be written in the
form
\begin{equation}\label{x.y}
\Lambda(\lambda) = \Lambda^{(1)}(\lambda) U(a) \Lambda^{(2)}(\lambda)
U(a)^{-1},
\end{equation}
where
\begin{equation*}
U(a)=\begin{pmatrix} e^{-i\sqrt{\lambda}a} & 0 \\ 0 & e^{i\sqrt{\lambda}a}
\end{pmatrix}.
\end{equation*}
The relation \eqref{x.y} is the special case of the well-known factorization
formula
\cite{Aktosun,Redheffer:61,Redheffer:62,Kowal,Rozman,Bianchi:Ventra:95,Bianchi:Ventra:95a}
applied to the Laplacian on a line with point interaction.

It is easy to realize that the transfer matrix cannot be defined for
\emph{arbitrary} boundary conditions. For instance, the Dirichlet
($\psi_2(0+)=\psi_1(0+)=0$) or Neuman ($\psi'_2(0+)=\psi'_1(0+)=0$) or mixed
($\psi_2(0+)+k_2\psi'_2(0+)=\psi_1(0+)+k_1\psi'_1(0+)=0$) boundary conditions
introduce the decoupling $\Delta(A,B)=\Delta_1\oplus \Delta_2$, where
$\Delta_j$, $j=1,2$ are the Laplacians on $L^2(0,\infty)$ with corresponding
boundary conditions. Recall, however, that the scattering matrix is well
defined even in these cases. The composition rule \eqref{spec.case} (see
Example \ref{graph:ex:5}) remains valid.

Now we consider an arbitrary graph $\Gamma$ with an even number of external
lines $n=2p$. We enumerate the external lines in an arbitrary but fixed order.
The external part of an arbitrary solution of the Schr\"{o}dinger equation with
$-\Delta(A,B)$ at the energy $\lambda>0$ has the form
\begin{equation}\label{graph:form}
u_j(x)=a_j e^{i\sqrt{\lambda}x} + b_j e^{-i\sqrt{\lambda}x},\qquad j=1,\ldots,
n.
\end{equation}
We define the transfer matrix
\begin{equation}\label{graph:transfer:matrix}
\Lambda(\lambda)\begin{pmatrix}
a_1 \\ \vdots \\ a_p \\ b_1 \\ \vdots \\ b_p
\end{pmatrix}=
\begin{pmatrix}
b_{p+1}\\ \vdots \\ b_{n} \\ a_{p+1} \\ \vdots \\ a_n
\end{pmatrix}.
\end{equation}
To prove that $\Lambda(\lambda)$ is correctly defined it suffices to show that
for arbitrary constants $(a_j, b_j)$, $j=1,\ldots,p$ there is a solution to the
Schr\"{o}dinger equation with the operator $-\Delta(A,B)$ whose external part has
the form \eqref{graph:form} and this solution is unique up to its internal
part. The external part of any solution to the Schr\"{o}dinger equation with the
operator $-\Delta(A,B)$ is a linear combination of the columns of the
matrix-valued function
\begin{equation}\label{xx.yy}
\Psi(x,\lambda)=e^{-i\sqrt{\lambda}x}\1+e^{i\sqrt{\lambda}x} S(\lambda).
\end{equation}
Thus, the columns of \eqref{xx.yy} have to satisfy
\eqref{graph:transfer:matrix}, i.e.
\begin{displaymath}
\Lambda(\lambda)\begin{pmatrix} S_{11}(\lambda) & S_{12}(\lambda) \\
\1 & 0 \end{pmatrix} = \begin{pmatrix} 0 & \1 \\
S_{21}(\lambda) & S_{22}(\lambda) \end{pmatrix},
\end{displaymath}
where the $p\times p$ block notation is adopted. Writing $\Lambda(\lambda)$ as
\begin{displaymath}
\Lambda(\lambda)=\begin{pmatrix} \Lambda_{11}(\lambda) & \Lambda_{12}(\lambda) \\
\Lambda_{21}(\lambda) & \Lambda_{22}(\lambda) \end{pmatrix}
\end{displaymath}
we obtain
\begin{equation}\label{wichtig}
\begin{split}
&\Lambda_{11}(\lambda)S_{11}(\lambda)+\Lambda_{12}(\lambda)=0,\\
&\Lambda_{11}(\lambda)S_{12}(\lambda)=\1,\\
&\Lambda_{21}(\lambda)S_{11}(\lambda)+\Lambda_{22}(\lambda) =S_{21}(\lambda),\\
&\Lambda_{21}(\lambda)S_{12}(\lambda) =S_{22}(\lambda).
\end{split}
\end{equation}

Let us suppose that $\det S_{12}(\lambda)\neq 0$. Then
\begin{eqnarray*}
\Lambda_{11}(\lambda)=S_{12}(\lambda)^{-1}, &&
\Lambda_{12}(\lambda)= - S_{12}(\lambda)^{-1} S_{11}(\lambda),\\
\Lambda_{21}(\lambda) = S_{22}(\lambda)
S_{12}(\lambda)^{-1}, && \Lambda_{22}(\lambda) = S_{21}(\lambda)-
S_{22}(\lambda)S_{12}(\lambda)^{-1}S_{11}(\lambda).
\end{eqnarray*}
Thus, we proved that for $\det S_{12}(\lambda)\neq 0$ the transfer matrix
exists and has the form
\begin{equation}\label{Lambda.form}
\Lambda(\lambda)=\begin{pmatrix}
S_{12}(\lambda)^{-1} & -S_{12}(\lambda)^{-1} S_{11}(\lambda)\\ S_{22}(\lambda)
S_{12}(\lambda)^{-1} & S_{21}(\lambda)-
S_{22}(\lambda)S_{12}(\lambda)^{-1}S_{11}(\lambda)
\end{pmatrix}.
\end{equation}
Also, its definition \eqref{graph:transfer:matrix} immediately leads to the
following factorization formula
\begin{equation}\label{comp:rule:alt}
\Lambda(\lambda)=\Lambda^{(1)}(\lambda)U(\underline{a})\Lambda^{(2)}(\lambda)
U(\underline{a})^{-1},
\end{equation}
where the diagonal unitary matrix $U(\underline{a})$ is given by
\begin{equation*}
U(\underline{a})=\begin{pmatrix} e^{-i\sqrt{\lambda}\underline{a}} & 0 \\ 0 &
e^{i\sqrt{\lambda}\underline{a}}
\end{pmatrix}.
\end{equation*}
Note that formal arguments based on the superposition principle leading to
\eqref{comp:rule:alt} have appeared earlier in \cite{Datta}.
As for related results we mention that in \cite{Kurasov:Pavlov} it was shown
that the transfer matrix of a Schr\"{o}dinger operator on the line with a
matrix-valued potential can be written in the form
\eqref{Lambda.form}.

\begin{lemma}
If $\det S_{12}(\lambda)\neq 0$ then $\Lambda(\lambda)\in\U(p,p)$.
\end{lemma}

\begin{proof}
Obviously the coefficients $a_1,\ldots,a_n,b_1,\ldots,b_n$ in
\eqref{graph:form} satisfy the relation
\begin{equation*}
\begin{pmatrix} a_1 \\ \vdots \\ a_n \end{pmatrix} = S(\lambda)
\begin{pmatrix} b_1 \\ \vdots \\ b_n \end{pmatrix}.
\end{equation*}
From the unitarity of the scattering matrix it follows that
\begin{equation*}
|a_1|^2+\ldots+|a_n|^2=|b_1|^2+\ldots+|b_n|^2,
\end{equation*}
or, equivalently,
\begin{equation*}
|a_1|^2+\ldots+|a_p|^2-|b_1|^2-|b_p|^2=|b_{p+1}|^2+\ldots+|b_n|^2
-|a_{p+1}|^2-\ldots-|a_n|^2.
\end{equation*}
This relation and \eqref{graph:transfer:matrix} complete the proof of the
lemma.
\end{proof}

Let us summarize the above results of the present Section:

\begin{theorem}\label{summarize}
If $\det S_{12}(\lambda)\neq 0$ then the transfer matrix
$\Lambda(\lambda)\in\U(p,p)$ as given by \eqref{Lambda.form} exists such that
for an arbitrary $(a_1,\ldots,a_p,b_1,\ldots,b_p)\in\C^n$ there is a solution
of the Schr\"{o}dinger equation with $-\Delta(A,B)$ at the energy $\lambda>0$ whose
external part has the form \eqref{graph:form} and the coefficients
$(b_{p+1},\ldots,b_n,a_{p+1},\ldots,a_n)\in\C^n$ are given by
\eqref{graph:transfer:matrix}. The composition rule for the scattering matrices
\eqref{compos} is equivalent to the multiplication formula \eqref{comp:rule:alt}
for the transfer matrices.
\end{theorem}

In addition for real operators we have

\begin{theorem}\label{real}
If the operator $\Delta(A,B,\underline{a})$ is real and if in addition $\det
S_{12}(\lambda)\neq 0$ then $\Lambda(\lambda)\in\SU(p,p)\subset\SL(2p;\C)$.
\end{theorem}

\begin{proof}
From the well-known determinant formula for block matrices (see e.g.\
\cite[Section II.5]{Gantmacher})
\begin{equation}\label{Gant}
\det\begin{pmatrix} A_{11} & A_{12} \\ A_{21} & A_{22} \end{pmatrix} = \det A_{11}\
\det\big(A_{22}-A_{21} A_{11}^{-1}A_{12} \big),
\end{equation}
which follows from the decomposition
\begin{equation*}
\begin{pmatrix} A_{11} & A_{12} \\ A_{21} & A_{22} \end{pmatrix}=
\begin{pmatrix} A_{11} & 0 \\ A_{21} & \1 \end{pmatrix}\
\begin{pmatrix} \1 & A_{11}^{-1}A_{12} \\ 0 & A_{22}-A_{21} A_{11}^{-1}A_{12} \end{pmatrix},
\end{equation*}
it follows that
\begin{equation*}
\det\Lambda(\lambda)=\frac{\det S_{21}(\lambda)}{\det S_{12}(\lambda)}.
\end{equation*}
By Theorem \ref{transp} we have $S_{12}(\lambda)^T=S_{21}(\lambda)$ and thus
$\det\Lambda(\lambda)=1$.
\end{proof}

We turn now to a discussion of the assumption $\det S_{12}(\lambda)\neq 0$. For
the scattering matrix of the graph depicted in Fig.\ \ref{graph:fig:1} with the
boundary conditions \eqref{randbed} (see Example \ref{graph:ex:6}) $\det
\left(S_a(\lambda)\right)_{12}=0$ for all $\lambda>0$.

\begin{theorem}\label{det.S.0}
Suppose that $\det S_{12}(\lambda)=0$. Then the transfer matrix
$\Lambda(\lambda)$ exists such that for arbitrary
\begin{displaymath}
(a_1,\ldots,a_p,b_1,\ldots,b_p)\in\Ran \begin{pmatrix} \1 & 0 \\ 0 & P_{(\Ker
S_{12}(\lambda))^\perp}
\end{pmatrix}\subset \C^n
\end{displaymath}
there is a solution of the Schr\"{o}dinger equation with
$-\Delta(A,B,\underline{a})$ at energy $\lambda>0$ whose external part has the
form \eqref{graph:form} and the coefficients
$(b_{p+1},\ldots,b_n,a_{p+1},\ldots,a_n)\in\C^n$ are given by
\eqref{graph:transfer:matrix}.
\end{theorem}

\begin{proof}
The external part of any solution to the Schr\"{o}dinger equation with the operator
$-\Delta(A,B,\underline{a})$ satisfying the conditions of the theorem is a
linear combination of the columns of the matrix-valued function
\begin{equation}\label{y.z}
\Psi(x,\lambda)\begin{pmatrix} \1 & 0 \\ 0 & P_{(\Ker
S_{12}(\lambda))^\perp}  \end{pmatrix},
\end{equation}
where $\Psi(x,\lambda)$ is given by \eqref{xx.yy}. Thus, the columns of
\eqref{y.z} have to satisfy \eqref{graph:transfer:matrix}, i.e.
\begin{equation*}
\begin{pmatrix} \Lambda_{11}(\lambda) & \Lambda_{12}(\lambda) \\
\Lambda_{21}(\lambda) & \Lambda_{22}(\lambda) \end{pmatrix}
\begin{pmatrix} S_{11}(\lambda) & S_{12}(\lambda)P_{(\Ker
S_{12}(\lambda))^\perp}\\ \1 & 0 \end{pmatrix}=
\begin{pmatrix} 0 &  P_{(\Ker
S_{12}(\lambda))^\perp} \\ S_{21}(\lambda) & S_{22}(\lambda)P_{(\Ker
S_{12}(\lambda))^\perp}\end{pmatrix}.
\end{equation*}
The solution of this equation can be wiritten in the form
\begin{eqnarray*}
\Lambda_{11}(\lambda)=S_{12}(\lambda)^\star,&& \Lambda_{12}(\lambda)=
- S_{12}(\lambda)^\star S_{11}(\lambda),\\
\Lambda_{21}(\lambda)= S_{22}(\lambda) S_{12}(\lambda)^\star,&&
\Lambda_{22}=S_{21}(\lambda)-S_{22}(\lambda) S_{12}(\lambda)^\star S_{11}(\lambda),
\end{eqnarray*}
where $\star$ stands for the Penrose-Moore pseudoinverse.
\end{proof}

Note that any vector of the form $(c,0)^T$ with $c\in\Ker S_{12}(\lambda)$
satisfies $\Lambda(\lambda) \begin{pmatrix} c\\ 0 \end{pmatrix}=0$. Thus
$\det\Lambda(\lambda)=0$.

Inspection of the proof of Theorem \ref{det.S.0} shows that the transfer matrix
cannot be extended to a subspace larger than
\begin{displaymath}
\Ran\begin{pmatrix} \1 & 0 \\ 0 & P_{(\Ker
S_{12}(\lambda))^\perp}  \end{pmatrix}.
\end{displaymath}
If $\det \Lambda^{(1)}(\lambda)=\det \Lambda^{(2)}(\lambda)=0$ then $\Ran
U(\underline{a})\Lambda^{(2)}(\lambda)U(\underline{a})$ and $\Ker
\Lambda^{(1)}(\lambda)$ may have a nontrivial overlap and therefore the
multiplication formula \eqref{comp:rule:alt} does not hold in this case.

\begin{example}
Consider the graph depicted in Fig.\ \ref{graph:fig:1} with the boundary
conditions from Example \ref{graph:ex:6}. For all $\lambda\in\R_+$ such that
$e^{2i\sqrt{\lambda}a}=1$ we have that $\Ker S_{12}(\lambda)$ is nontrivial and
\begin{eqnarray*}
P_{\Ker S_{12}(\lambda)}=\frac{1}{2}\begin{pmatrix} 1 & \mp 1 \\
\mp 1 & 1 \end{pmatrix},\\
P_{(\Ker S_{12}(\lambda))^\perp}=\frac{1}{2}\begin{pmatrix} 1 & \pm 1 \\
\pm 1 & 1 \end{pmatrix},
\end{eqnarray*}
where $\pm 1$ corresponds to $\exp\{i\sqrt{\lambda}a\}=\pm 1$. Suppose that
\begin{equation*}
(a_1,a_2,b_1,b_2)^T\notin\Ran\begin{pmatrix}\1 & 0 \\ 0 & P_{(\Ker
S_{12}(\lambda))^\perp} \end{pmatrix},
\end{equation*}
or, equivalently,
\begin{equation*}
(a_1,a_2,b_1,b_2)^T\in\Ran\begin{pmatrix}0 & 0 \\ 0 & P_{\Ker S_{12}(\lambda)}
\end{pmatrix}.
\end{equation*}
In particular, we can choose
\begin{equation*}
a_1=a_2=0,\qquad b_1=1,\qquad b_2=\mp 1.
\end{equation*}
It is easy to check that there is no solution to the Schr\"{o}dinger equation with
these boundary conditions.
\end{example}

\begin{example}
Consider the graph depicted in Fig.\ \ref{graph:fig:1a} with the boundary
conditions as in Example \ref{graph:ex:7}. $\Ker S_{12}(\lambda)$ is nontrivial
for all $\lambda\in\R_+$ and
\begin{equation*}
P_{\Ker S_{12}(\lambda)} =\frac{1}{2}\begin{pmatrix} 1 & -1 \\ -1 &
1\end{pmatrix},\qquad P_{(\Ker S_{12}(\lambda))^\perp}
=\frac{1}{2}\begin{pmatrix} 1 &
1 \\ 1 & 1\end{pmatrix}.
\end{equation*}
Suppose again that
\begin{equation*}
(a_1,a_2,b_1,b_2)^T\notin\Ran\begin{pmatrix}\1 & 0 \\ 0 & P_{(\Ker
S_{12}(\lambda))^\perp} \end{pmatrix}
\end{equation*}
and choose
\begin{equation*}
a_1=a_2=0,\qquad b_1=1,\qquad b_2=1.
\end{equation*}
Again it is easy to check that there is no solution to the Schr\"{o}dinger equation
with these boundary conditions.
\end{example}

The statement converse to Theorem \ref{summarize} immediately follows from
Theorem \ref{det.S.0}.

\begin{theorem}
If the transfer matrix $\Lambda(\lambda)$ exists in the sense of Theorem
\ref{summarize} then $\det S_{12}(\lambda)\neq 0$ and the corresponding
scattering matrix is given by
\begin{equation}\label{S.repres}
S(\lambda)=\begin{pmatrix} -\Lambda_{11}(\lambda)^{-1}\Lambda_{12}(\lambda) &
\Lambda_{11}(\lambda)^{-1}\\
\Lambda_{22}(\lambda)-\Lambda_{21}(\lambda)\Lambda_{11}(\lambda)^{-1} \Lambda_{12}(\lambda) &
\Lambda_{21}(\lambda)\Lambda_{11}(\lambda)^{-1}
\end{pmatrix}.
\end{equation}
\end{theorem}

\begin{proof}
Suppose that $\det S_{12}(\lambda)=0$. Then by Theorem \ref{det.S.0} we get
$\det \Lambda(\lambda)=0$, which is a contradiction. Thus, $\det
S_{12}(\lambda)\neq 0$ and therefore by Theorem \ref{summarize} $\det
\Lambda(\lambda)\neq 0$. The representation \eqref{S.repres} follows from
\eqref{wichtig}.
\end{proof}

\section*{Appendix A}
\setcounter{equation}{0}
\renewcommand{\theequation}{A.\arabic{equation}}

Here we give the proof of Theorem 3.6 which claims that for arbitrary unitary
matrices $U^{(1)}$, $U^{(2)}$, and $V$ the matrix $U=U^{(1)}*_V U^{(2)}$
defined by \eqref{56} is unitary. As already noted in
\cite{Kostrykin:Schrader:99b} it suffices to prove only the relations
\begin{equation}\label{app.1}
\begin{split}
{U_{11}}^\ast U_{11} + {U_{21}}^\ast U_{21} & =\1,\\ {U_{11}}^\ast U_{12} +
{U_{21}}^\ast U_{22} & = 0.
\end{split}
\end{equation}
The remaining relations
\begin{equation}\label{app.1.Extra}
\begin{split}
{U_{12}}^\ast U_{12} + {U_{22}}^\ast U_{22} & =\1,\\ {U_{12}}^\ast U_{11} +
{U_{22}}^\ast U_{21} & = 0.
\end{split}
\end{equation}
follow immediately from \eqref{app.1}. To see this for an arbitrary unitary
matrix $U$ we define an involutive map $U\mapsto U^\tau$ given as
\begin{equation*}
U=\begin{pmatrix} U_{11} & U_{12} \\ U_{21} & U_{22} \end{pmatrix}\quad
\mapsto\quad U^\tau = \begin{pmatrix} U_{22} & U_{21} \\ U_{12} & U_{11} \end{pmatrix}.
\end{equation*}
Direct calculations show that the following ``transposition law"
\begin{equation}\label{u.tau.Def}
U^\tau = {U^{(2)}}^\tau *_{V^\ast} {U^{(1)}}^\tau
\end{equation}
holds whenever $U=U^{(1)} *_V U^{(2)}$. Assume that \eqref{app.1} holds for
arbitrary unitary $U$. Replacing the matrix $U$ by $U^\tau$ given by
\eqref{u.tau.Def} transforms the relations \eqref{app.1} into \eqref{app.1.Extra}.

By the definition of the generalized star product \eqref{56} and by the
unitarity of $U^{(1)}$ the first of the relations \eqref{app.1} is equivalent
to
\begin{equation}\label{app.2}
\begin{split}
-&{U_{21}^{(1)}}^\ast U_{21}^{(1)} + {U_{21}^{(1)}}^\ast V^\ast
{U_{11}^{(2)}}^\ast K_2^\ast {U_{12}^{(1)}}^\ast U_{11}^{(1)}+
{U_{11}^{(1)}}^\ast U_{12}^{(1)} K_2 U_{11}^{(2)} V U_{21}^{(1)}\\ &+
{U_{21}^{(1)}}^\ast V^\ast {U_{11}^{(2)}}^\ast K_2^\ast {U_{12}^{(1)}}^\ast
U_{12}^{(1)} K_2 U_{11}^{(2)} V U_{21}^{(1)}+ {U_{21}^{(1)}}^\ast K_1^\ast
{U_{21}^{(2)}}^\ast U_{21}^{(2)} K_1 U_{21}^{(1)} = 0.
\end{split}
\end{equation}
Since the opposite case was already considered in \cite{Kostrykin:Schrader:99b}
we further assume that the matrix $U^{(1)}$ is not $V$-compatible with
$U^{(2)}$. From Theorem \ref{graph:dosta} it follows that all off-diagonal
blocks $U_{12}^{(1)}$, $U_{21}^{(1)}$, $U_{12}^{(2)}$, and $U_{21}^{(2)}$ are
not of maximal rank and thus $\Ker U_{21}^{(1)}$ is nontrivial. Let $d_i$,
$1\leq i\leq k=\dim \Ker U_{21}^{(1)}$ be an arbitrary basis in $\Ker
U_{21}^{(1)}$. From the unitarity of the matrix $U^{(1)}$ we get
${U_{12}^{(1)}}^\ast U_{11}^{(1)} d_i = 0$ for all $1\leq i\leq k$. Thus
\begin{equation*}
\begin{split}
&\Big[-{U_{21}^{(1)}}^\ast U_{21}^{(1)} + {U_{21}^{(1)}}^\ast V^\ast
{U_{11}^{(2)}}^\ast K_2^\ast {U_{12}^{(1)}}^\ast U_{11}^{(1)}+
{U_{11}^{(1)}}^\ast U_{12}^{(1)} K_2 U_{11}^{(2)} V U_{21}^{(1)}\\ &+
{U_{21}^{(1)}}^\ast V^\ast {U_{11}^{(2)}}^\ast K_2^\ast {U_{12}^{(1)}}^\ast
U_{12}^{(1)} K_2 U_{11}^{(2)} V U_{21}^{(1)}+ {U_{21}^{(1)}}^\ast K_1^\ast
{U_{21}^{(2)}}^\ast U_{21}^{(2)} K_1 U_{21}^{(1)}\Big] d_i = 0
\end{split}
\end{equation*}
for all $1\leq i\leq k$. Hence to prove \eqref{app.2} it remains to show that
\begin{equation}\label{app.3}
\begin{split}
&\Big[-{U_{21}^{(1)}}^\ast U_{21}^{(1)} + {U_{21}^{(1)}}^\ast V^\ast
{U_{11}^{(2)}}^\ast K_2^\ast {U_{12}^{(1)}}^\ast U_{11}^{(1)}+
{U_{11}^{(1)}}^\ast U_{12}^{(1)} K_2 U_{11}^{(2)} V U_{21}^{(1)}\\ &+
{U_{21}^{(1)}}^\ast V^\ast {U_{11}^{(2)}}^\ast K_2^\ast {U_{12}^{(1)}}^\ast
U_{12}^{(1)} K_2 U_{11}^{(2)} V U_{21}^{(1)}+ {U_{21}^{(1)}}^\ast K_1^\ast
{U_{21}^{(2)}}^\ast U_{21}^{(2)} K_1 U_{21}^{(1)}\Big] d = 0
\end{split}
\end{equation}
for any $d\in\left(\Ker U_{21}^{(1)}\right)^\perp =\Ran {U_{21}^{(1)}}^\ast$.
Therefore we set $d={U_{21}^{(1)}}^\ast \widetilde{d}$, where
$\widetilde{d}\in\C^p$ is an arbitrary vector. Thus, the relation \eqref{app.3}
holds whenever
\begin{equation}\label{A.5A}
\begin{split}
&\Big[-{U_{21}^{(1)}}^\ast U_{21}^{(1)}{U_{21}^{(1)}}^\ast +
{U_{21}^{(1)}}^\ast V^\ast {U_{11}^{(2)}}^\ast K_2^\ast {U_{12}^{(1)}}^\ast
U_{11}^{(1)}{U_{21}^{(1)}}^\ast+ {U_{11}^{(1)}}^\ast U_{12}^{(1)} K_2
U_{11}^{(2)} V U_{21}^{(1)}{U_{21}^{(1)}}^\ast\\ &+ {U_{21}^{(1)}}^\ast V^\ast
{U_{11}^{(2)}}^\ast K_2^\ast {U_{12}^{(1)}}^\ast U_{12}^{(1)} K_2 U_{11}^{(2)}
V U_{21}^{(1)}{U_{21}^{(1)}}^\ast+ {U_{21}^{(1)}}^\ast K_1^\ast
{U_{21}^{(2)}}^\ast U_{21}^{(2)} K_1 U_{21}^{(1)}{U_{21}^{(1)}}^\ast\Big]
\widetilde{d}= 0
\end{split}
\end{equation}
for all $\widetilde{d}\in\C^p$.

First we note that by Lemma \ref{graph:neu:i:v} (iv) the relation \eqref{A.5A}
holds for all $\widetilde{d}\in\widetilde{\cC}$. Therefore it suffices to prove
that \eqref{A.5A} holds for all $\widetilde{d}\in\widetilde{\cC}^\perp$.
Observe that in this case by Lemma \ref{graph:neu:i:v} (i) and by the unitarity
of the matrices $U^{(1)}$ and $U^{(2)}$ we have
\begin{equation}\label{app.incl.neu.1}
\begin{split}
{U_{12}^{(1)}}^\ast U_{11}^{(1)} {U_{21}^{(1)}}^\ast \widetilde{d} &= -
{U_{12}^{(1)}}^\ast U_{12}^{(1)} {U_{22}^{(1)}}^\ast \widetilde{d} =
-{U_{22}^{(1)}}^\ast \widetilde{d} + {U_{22}^{(1)}}^\ast U_{22}^{(1)} {U_{22}^{(1)}}^\ast
\widetilde{d} \in \cB^\perp,\\
U_{11}^{(2)} V U_{21}^{(1)} {U_{21}^{(1)}}^\ast \widetilde{d} & = U_{11}^{(2)}
 V \widetilde{d} - U_{11}^{(2)} V U_{22}^{(1)} {U_{22}^{(1)}}^\ast \widetilde{d}
 \in \widetilde{\cB}^\perp,\\
U_{21}^{(1)} {U_{21}^{(1)}}^\ast \widetilde{d} &= \widetilde{d} - U_{22}^{(1)}
{U_{22}^{(1)}}^\ast \widetilde{d} \in \widetilde{\cC}^\perp.
\end{split}
\end{equation}
To prove the first relation in \eqref{app.incl.neu.1} it suffices to show that
for any $\widetilde{d}\in\widetilde{\cC}^\perp$ and any $b\in\cB$
\begin{equation}\label{Ziel.neu.1}
-  (b, {U_{22}^{(1)}}^\ast\widetilde{d})+(b, {U_{22}^{(1)}}^\ast U_{22}^{(1)}
{U_{22}^{(1)}}^\ast \widetilde{d}) = -(U_{22}^{(1)} b, \widetilde{d})+
(U_{22}^{(1)} {U_{22}^{(1)}}^\ast U_{22}^{(1)}b, \widetilde{d}) = 0.
\end{equation}
By the definition of $\cB$ and by Lemma \ref{graph:lem:factor}
${U_{22}^{(1)}}^\ast U_{22}^{(1)} b = b$ for any $b\in\cB$ which proves
\eqref{Ziel.neu.1}. To prove the second relation in \eqref{app.incl.neu.1}
it suffices to show for any $\widetilde{d}\in\widetilde{\cC}^\perp$ and any
$\widetilde{b}\in\widetilde{\cB}$
\begin{equation}\label{Ziel.neu.2}
\begin{split}
& (\widetilde{b}, U_{11}^{(2)}V \widetilde{d}) - (\widetilde{b}, U_{11}^{(2)} V
U_{22}^{(1)} {U_{22}^{(1)}}^\ast \widetilde{d})
=\\ & (V^\ast U_{11}^{(2)} \widetilde{b}, \widetilde{d})
- (U_{22}^{(1)}{U_{22}^{(1)}}^\ast V^\ast {U_{11}^{(2)}}^\ast\widetilde{b}, \widetilde{d})
= 0.
\end{split}
\end{equation}
By Lemma \ref{graph:neu:i:v} (i) $V^\ast
U_{11}^{(2)}\widetilde{b}\in\widetilde{\cC}$. By the definition of
$\widetilde{\cC}$ and by Lemma \ref{graph:lem:factor} $U_{22}^{(1)}
{U_{22}^{(1)}}^\ast \widetilde{c} = \widetilde{c}$ for any
$\widetilde{c}\in\widetilde{\cC}$ which proves \eqref{Ziel.neu.2}. This also
proves that
\begin{equation*}
(\widetilde{c}, \widetilde{d}) - (\widetilde{c}, U_{22}^{(1)}
{U_{22}^{(1)}}^\ast \widetilde{d}) = (\widetilde{c}, \widetilde{d}) -
(U_{22}^{(1)} {U_{22}^{(1)}}^\ast \widetilde{c}, \widetilde{d}) = 0
\end{equation*}
for all $\widetilde{d}\in\widetilde{\cC}^\perp$ and all
$\widetilde{c}\in\widetilde{\cC}$ from which the third relation in
\eqref{app.incl.neu.1} follows.

From Lemma \ref{X.X.neu} and the definition \eqref{K.def} of the matrices $K_1$
and $K_2$ it follows that
\begin{equation}\label{app.incl.neu.2}
\begin{split}
\textrm{(i)}\quad & K_1\ \textrm{maps}\ \widetilde{\cC}^\perp\ \textrm{onto}\ \cC^\perp\
\textrm{bijectively},\\ \textrm{(ii)}\quad & K_1^\ast\ \textrm{maps}\ \cC^\perp\
\textrm{onto}\ \widetilde{\cC}^\perp\
\textrm{bijectively},\\ \textrm{(iii)}\quad & K_2\ \textrm{maps}\ \widetilde{\cB}^\perp\
\textrm{onto}\ \cB^\perp\
\textrm{bijectively},\\ \textrm{(iv)}\quad & K_2^\ast\ \textrm{maps}\ \cB^\perp\
\textrm{onto}\ \widetilde{\cB}^\perp\
\textrm{bijectively}.
\end{split}
\end{equation}

Noting that ${U_{12}^{(1)}}^\ast U_{12}^{(1)} b_\perp = b_\perp-
{U_{22}^{(1)}}^\ast U_{22}^{(1)} b_\perp \in \cB^\perp$ and
${U_{21}^{(2)}}^\ast U_{21}^{(2)} c_\perp = c_\perp - {U_{11}^{(2)}}^\ast
U_{11}^{(2)} c_\perp \in\cC^\perp$ due to \eqref{app.incl.neu.1} and
\eqref{app.incl.neu.2} we can write the l.h.s.\ of \eqref{A.5A} in the form
\begin{equation}\label{A.5B}
\begin{split}
&{U_{21}^{(1)}}^\ast\big[ -\1 - V^\ast {U_{11}^{(2)}}^\ast K_2
{U_{22}^{(1)}}^\ast - U_{22}^{(1)} K_2 U_{11}^{(2)} V + V^\ast
{U_{11}^{(2)}}^\ast K_2^\ast K_2 U_{11}^{(2)} V \\ & - V^\ast
{U_{11}^{(2)}}^\ast K_2^\ast {U_{22}^{(1)}}^\ast U_{22}^{(1)} K_2 U_{11}^{(2)}
V + K_1^\ast K_1 - K_1^\ast {U_{11}^{(2)}}^\ast U_{11}^{(2)} K_1 \big]
{U_{21}^{(2)}}^\ast U_{21}^{(2)} \widetilde{d}.
\end{split}
\end{equation}
Similar to \cite{Kostrykin:Schrader:99b} one can easily prove that for any
$\widetilde{b}_\perp\in\widetilde{\cB}^\perp$ and
$\widetilde{c}_\perp\in\widetilde{\cC}^\perp$ the following relations hold:
\begin{eqnarray*}
\lefteqn{K_1^\ast K_1 \widetilde{c}_\perp = \widetilde{c}_\perp + U_{22}^{(1)} K_2
U_{11}^{(2)} V \widetilde{c}_\perp + V^\ast {U_{11}^{(2)}}^\ast K_2^\ast
{U_{22}^{(1)}}^\ast \widetilde{c}_\perp}\\ &&+ V^\ast {U_{11}^{(2)}}^\ast
K_2^\ast {U_{22}^{(1)}}^\ast U_{22}^{(1)} K_2 U_{11}^{(2)} V
\widetilde{c}_\perp,
\end{eqnarray*}
\begin{eqnarray*}
K_1^\ast {U_{11}^{(2)}}^\ast U_{11}^{(2)} K_1 \widetilde{c}_\perp = V^\ast (\1
+ {U_{11}^{(2)}}^\ast K_2^\ast {U_{22}^{(1)}}^\ast V^\ast) {U_{11}^{(2)}}^\ast
U_{11}^{(2)}(\1 + V U_{22}^{(1)} K_2 U_{11}^{(2)})V\widetilde{c}_\perp,
\end{eqnarray*}
\begin{eqnarray*}
\lefteqn{K_2^\ast K_2 \widetilde{b}_\perp = \widetilde{b}_\perp + K_2^\ast
{U_{22}^{(1)}}^\ast V^\ast {U_{11}^{(2)}}^\ast + U_{11}^{(2)} V U_{22}^{(1)}
K_2 \widetilde{b}_\perp}\\ && + K_2^\ast {U_{22}^{(1)}}^\ast V^\ast
{U_{11}^{(2)}}^\ast U_{11}^{(2)} V U_{22}^{(1)} K_2 \widetilde{b}_\perp.
\end{eqnarray*}
Inserting these relations in \eqref{A.5B} with the choice $\widetilde{c}_\perp
= {U_{21}^{(2)}}^\ast U_{21}^{(2)} \widetilde{d}$ and
$\widetilde{b}_\perp = U_{11}^{(2)} V \widetilde{c}_\perp$ we obtain that it
vanishes thus completing the proof of the first relation in
\eqref{app.1}. The proof of the second relation in \eqref{app.1} is similar and
will therefore be omitted.

\end{document}